\documentclass[10pt,journal,compsoc]{IEEEtran}

\usepackage{amsmath,amssymb,amsthm}
\usepackage{graphicx}
\usepackage{colortbl}
\usepackage[ruled,vlined,linesnumbered]{algorithm2e}

\SetAlFnt{\small}
\SetAlCapFnt{\small}
\SetAlCapNameFnt{\small}
\SetAlCapHSkip{0pt}
\IncMargin{-\parindent}
\usepackage{xspace}
\usepackage[dvipsnames,x11names]{xcolor}
\usepackage{listings}
\usepackage{tabularx}
\usepackage{multirow}
\usepackage[inline,shortlabels]{enumitem}
\usepackage{fontawesome}
\usepackage{caption}
\usepackage{subcaption}
\usepackage{booktabs}
\usepackage[normal]{threeparttable}
\usepackage{wasysym}
\usepackage{verbatim}
\usepackage{microtype}
\usepackage[hyphens]{url}
\usepackage[hidelinks]{hyperref}
\usepackage{tikz}
\usetikzlibrary{trees,intersections,arrows}
\usepackage{tkz-euclide}
\usetkzobj{all}
\usepackage{shortcuts} %
\usepackage{drawstack}

\begin{document}

\title{The Art, Science, and Engineering of Fuzzing:\\A Survey}

\author{Valentin J.M. Man\`es,
HyungSeok Han,
Choongwoo Han,
Sang Kil Cha,
Manuel Egele,\\
Edward J. Schwartz,
and Maverick Woo%
\IEEEcompsocitemizethanks{
    \IEEEcompsocthanksitem V. J. M. Man\`es is with KAIST Cyber Security Research
    Center, Korea \protect\\ E-mail: valentin.manes@kaist.ac.kr.
    \IEEEcompsocthanksitem H. Han and S. K. Cha are with KAIST, Korea
    \protect\\ E-mail: hyungseok.han@kaist.ac.kr and sangkilc@kaist.ac.kr.
    \IEEEcompsocthanksitem C. Han is with Naver Corp., Korea \protect\\
    E-mail: cwhan.tunz@navercorp.com.
    \IEEEcompsocthanksitem M. Egele is with Boston University
    \protect\\ E-mail: megele@bu.edu.
    \IEEEcompsocthanksitem E. J. Schwartz is with SEI, Carnegie Mellon University
    \protect\\ E-mail: edmcman@cmu.edu
    \IEEEcompsocthanksitem M. Woo is with Carnegie Mellon
    University \protect\\ E-mail: pooh@cmu.edu.
}
\thanks{Corresponding author: Sang Kil Cha.}
\thanks{Manuscript submitted on April 8, 2019.}
}

\IEEEtitleabstractindextext{
    \begin{abstract}

        Among the many software vulnerability discovery techniques available
        today, \emph{fuzzing} has remained highly popular due to its
        conceptual simplicity, its low barrier to deployment, and its vast
        amount of empirical evidence in discovering real-world software
        vulnerabilities.
        At a high level, fuzzing refers to a process of repeatedly
        running a program with generated inputs that may be
        syntactically or semantically malformed.
        While researchers and practitioners alike have invested a large and
        diverse effort towards improving fuzzing in recent years, this surge
        of work has also made it difficult to gain a comprehensive and
        coherent view of fuzzing.
        To help preserve and bring coherence to the vast literature of
        fuzzing, this paper presents a unified, general-purpose model of
        fuzzing together with a taxonomy of
        the current fuzzing literature.
        We methodically explore the design decisions at every stage of our model
        fuzzer by surveying the related literature and innovations in the art,
        science, and engineering that make modern-day fuzzers effective.

    \end{abstract}

}

\maketitle

\section{Introduction}\label{sec:intro}

Ever since its introduction in the early 1990s~\cite{miller:cacm:1990},
\emph{fuzzing} has remained one of the most widely-deployed techniques to
discover software security vulnerabilities.
At a high level, fuzzing refers to a process of repeatedly running a program
with generated inputs that may be syntactically or semantically malformed.
In practice, attackers routinely deploy fuzzing in scenarios such as exploit
generation and penetration testing~\cite{arstechnica:pwn2own:fuzz,
  miller:fuzz:first}; several teams in the 2016 DARPA Cyber Grand Challenge
(CGC) also employed fuzzing in their cyber reasoning
systems~\cite{bohme:ccs:2016, stephens:ndss:2016, afl-used-by-fas,
  afl-used-by-grammatech}.
Fueled by these activities, defenders have started to use fuzzing in an attempt
to discover vulnerabilities before attackers do.
For example, prominent vendors such as Adobe~\cite{adobe:binspector},
Cisco~\cite{cisco:security-program}, Google~\cite{clusterfuzz,
google:chromium-security, aizatsky:ossfuzz:2016}, and
Microsoft~\cite{bounimova:icse:2013, microsoft:sdl:fuzzstep} all employ fuzzing
as part of their secure development practices.
More recently, security auditors~\cite{interceptfuzz} and open-source
developers~\cite{fuzzing-project:software} have also started to use fuzzing to
gauge the security of commodity software packages and provide some suitable
forms of assurance to end-users.

The fuzzing community is extremely vibrant.
As of this writing, GitHub alone hosts over a thousand public repositories
related to fuzzing~\cite{githubfuzzer}.
And as we will demonstrate, the literature also contains a large number of
fuzzers (see Figure~\ref{fig:genealogy} on p.~\pageref{fig:genealogy}) and an
increasing number of fuzzing studies appear at major security
conferences (e.g.~\cite{woo:ccs:2013, cha:oakland:2015, bohme:ccs:2016,
rawat:ndss:2017, gan:oakland:2018, zhao:ndss:2019}). %
\mav{Do we want to revise this list, which is a bit dated by now? Reviewers may
like to see papers that are more recent (and it also helps to reduce citing our
own papers).}
\val{Better now?}
In addition, the blogosphere is filled with many success stories of fuzzing,
some of which also contain what we consider to be gems that warrant a permanent
place in the literature.%

Unfortunately, this surge of work in fuzzing by researchers and practitioners
alike also bears a warning sign of impeded progress.
For example, the description of some fuzzers do not go much beyond their
source code and manual page.
As such, it is easy to lose track of the design decisions and potentially
important tweaks in these fuzzers over time.
Furthermore, there has been an observable fragmentation in the terminology used
by various fuzzers.
For example, whereas AFL~\cite{aflfuzz} uses the term ``test case minimization''
to refer to a technique that reduces the size of a crashing input, the same
technique is called ``test case reduction'' in funfuzz~\cite{funfuzz}.
At the same time, while BFF~\cite{bff-deprecated-merged-with-certfuzz} includes
a similar-sounding technique called ``crash minimization'', this technique
actually seeks to minimize the number of bits that differ between the crashing
input and the original seed file and is not related to reducing input size.
This makes it difficult, if not impossible, to compare fuzzers using the
published evaluation results.
We believe such fragmentation makes it difficult to discover and disseminate
fuzzing knowledge and this may severely hinder the progress in fuzzing research
in the long run.

Due to the above reasons, we believe it is prime time to consolidate and distill
the large amount of progress in fuzzing, many of which happened after the three
trade-books on the subject were published in 2007--2008~\cite{evron:2007,
sutton:2007, takanen:2008}.

As we attempt to unify the field, we will start by using~\S\ref{sec:def} to
present our fuzzing terminology and a unified model of fuzzing.
Staying true to the purpose of this paper, our terminology is chosen to closely
reflect the current predominant usages, and
our model fuzzer (Algorithm~\ref{algo:fuzzer}, p.~\pageref{algo:fuzzer}) is
designed to suit a large number of fuzzing tasks as classified in a taxonomy of
the current fuzzing literature (Figure~\ref{fig:genealogy},
p.~\pageref{fig:genealogy}).
With this setup, we will then explore every stage of our model
fuzzer in \S\ref{sec:preprocess}--\S\ref{sec:confupdate}, and present a
detailed overview of major fuzzers in Table~\ref{tab:overview}
(p.~\pageref{tab:overview}).
At each stage, we will survey the relevant literature to explain the design
choices, discuss important trade-offs, and highlight many marvelous engineering
efforts that help make modern-day fuzzers effective at their task.

\section{Systemization, Taxonomy, and Test Programs}\label{sec:def}

\ed{Do we still want to use systemization here?}
The term ``fuzz'' was originally coined by Miller \etal in 1990 to refer to a
program that ``generates a stream of random characters to be consumed by a
target program''~\cite[p.~4]{miller:cacm:1990}.
Since then, the concept of fuzz as well as its action---``fuzzing''---has
appeared in a wide variety of contexts,
including dynamic symbolic
execution~\cite{godefroid:ndss:2008,xie:dsn:2009}, grammar-based test case
generation~\cite{godefroid:pldi:2008,holler:usec:2012,veggalam:esorics:2016},
permission testing~\cite{au:ccs:2012,felt:ccs:2011}, behavioral
testing~\cite{kapravelos:usec:2014,rasthofer:icse:2017,wong:ndss:2016},
complexity testing~\cite{lemieux:issta:2018, wei:fse:2018},
kernel testing~\cite{syzkaller, song:ndss:2019, schwarz:asiaccs:2018},
representation dependence testing~\cite{kanade:fse:2010},
function detection~\cite{xu:oakland:2017}, robustness
evaluation~\cite{winter:icse:2011}, exploit development~\cite{jana:oakland:2012},
GUI testing~\cite{song:ase:2017},
signature generation~\cite{dolan-gavitt:ccs:2009}, and penetration
testing~\cite{webfuzzlist,mulliner:usec:2011}.
To systematize\ed{... and here?} the knowledge from the vast literature of fuzzing, let us first
present a terminology of fuzzing extracted from modern uses.

\subsection{Fuzzing \& Fuzz Testing}
\label{ssec:fuzzdef}

Intuitively, fuzzing is the action of running a Program Under Test (PUT) with
``fuzz inputs''. %
Honoring Miller \etal, we consider a fuzz input to be an input that the PUT
\emph{may not} be expecting, i.e., an input that the PUT may process incorrectly
and trigger a behavior that was unintended by the PUT developer.
To capture this idea, we define the term \emph{fuzzing} as follows.
\begin{definition}[\emph{Fuzzing}]
  Fuzzing is the execution of the PUT using input(s) sampled from an input space
  (the ``fuzz input space'') that \emph{protrudes} the expected input space of
  the PUT.
\end{definition}
Three remarks are in order.
First, although it may be common to see the fuzz input space to contain the
expected input space, this is \emph{not} necessary---it suffices for the former
to contain an input \emph{not in} the latter.
Second, in practice fuzzing almost surely runs for \emph{many} iterations; thus
writing ``repeated executions'' above would still be largely accurate.
Third, the sampling process is \emph{not} necessarily randomized, as we will see
in~\S\ref{sec:inputgen}.

\emph{Fuzz testing} is a form of software testing technique that utilizes
fuzzing.
To differentiate it from others and to honor what we consider to be its most
prominent purpose, we deem it to have a specific goal of finding
security-related bugs, which include program crashes.
In addition, we also define \emph{fuzzer} and \emph{fuzz campaign}, both of
which are common terms in fuzz testing:
\begin{definition}[\emph{Fuzz Testing}]
  Fuzz testing is the use of fuzzing to test if a PUT violates a
  security policy.
\end{definition}
\begin{definition}[\emph{Fuzzer}]
  A fuzzer is a program that performs fuzz testing on a PUT.
\end{definition}

\begin{definition}[\emph{Fuzz Campaign}]
  A fuzz campaign is a specific execution of a fuzzer on a PUT with a specific
  security policy.
\end{definition}

The goal of running a PUT through a fuzzing campaign is to find
bugs~\cite{avizienis:2004} that violate the specified security policy.
For example, a security policy employed by early fuzzers tested only whether a
generated input---the \emph{test case}---crashed the PUT.
However, fuzz testing can actually be used to test any security policy
observable from an execution, i.e., EM-enforceable~\cite{schneider:2000}.
The specific mechanism that decides whether an execution violates the security
policy is called the \emph{bug oracle}.
\begin{definition}[\emph{Bug Oracle}]
  A bug oracle is a program, perhaps as part of a fuzzer, that determines
  whether a given execution of the PUT violates a specific security policy.
\end{definition}

We refer to the algorithm implemented by a fuzzer simply as its ``fuzz
algorithm''.
Almost all fuzz algorithms depend on some parameters beyond (the path to) the
PUT.
Each concrete setting of the parameters is a \emph{fuzz configuration}:
\begin{definition}[\emph{Fuzz Configuration}]
  A fuzz configuration of a fuzz algorithm comprises the parameter value(s) that
  control(s) the fuzz algorithm.
\end{definition}

The definition of a fuzz configuration is intended to be broad.
Note that the type of values in a fuzz configuration depend on the type of the
fuzz algorithm.
For example, a fuzz algorithm that sends streams of random bytes to the
PUT~\cite{miller:cacm:1990} has a simple configuration space
$\{ (\text{PUT}) \}$.
On the other hand, sophisticated fuzzers contain algorithms that accept a set of
configurations and evolve the set over time---this includes adding and removing
configurations.
For example, CERT BFF~\cite{bff-deprecated-merged-with-certfuzz} varies both the
mutation ratio and the seed over
the course of a campaign, and thus its configuration space is
$\{ (\text{PUT}, s_1, r_1), (\text{PUT}, s_2, r_2), \ldots \}$.
A seed is a (commonly well-structured) input to the PUT, used to generate test
cases by modifying it.
Fuzzers typically maintain a collection of seeds, and some fuzzers
evolve the collection as the fuzz campaign progresses. This collection
is called a \emph{seed pool}.
Finally, a fuzzer is able to store some data within each configuration.
For instance, coverage-guided fuzzers may store the attained coverage in each
configuration.

\subsection{Paper Selection Criteria}
\label{ssec:paper-select}
To achieve a well-defined scope, we have chosen to include all publications on
fuzzing in the last proceedings of 4 major security conferences and 3 major
software engineering conferences from Jan 2008 to February 2019.
Alphabetically, the former includes
\begin{enumerate*}[(i)]
\item ACM Conference on Computer and Communications Security (CCS),
\item IEEE Symposium on Security and Privacy (S\&P),
\item Network and Distributed System Security Symposium (NDSS), and
\item USENIX Security Symposium (USEC);
\end{enumerate*}
and the latter includes
\begin{enumerate*}[(i)]
\item ACM International Symposium on the Foundations of Software Engineering
  (FSE),
\item IEEE/ACM International Conference on Automated Software Engineering (ASE),
  and
\item International Conference on Software Engineering (ICSE).
\end{enumerate*}
For writings that appear in other venues or mediums, we include them based on
our own judgment on their relevance.

As mentioned in \S\ref{ssec:fuzzdef}, \emph{fuzz testing}
only differentiates itself from software testing in that fuzz testing
is security related.
In theory, focusing on security bugs does not imply a difference in
the testing process beyond the selection of a bug oracle. The
techniques used often vary in practice, however.
When designing a testing tool, access to source code and some knowledge about
the PUT are often assumed.
Such
assumptions often drive the development of testing tools to have
different characteristics compared to those of fuzzers, which are more
likely to be employed by parties other than the PUT's developer.
Nevertheless, the two fields are still closely related to one
another. Therefore, when we are unsure whether to classify a
publication as relating to ``fuzz testing'' and include it in this
survey, we follow a simple rule of thumb: we include the publication
if the word \emph{fuzz} appears in it.

\subsection{Fuzz Testing Algorithm}
\label{sec:fuzz-algo}

\begin{algorithm}[t]
\small
\DontPrintSemicolon
\SetKwSty{algokeywordsty}
\SetFuncSty{algofuncsty}
\SetDataSty{algodatasty}
\SetArgSty{algoargsty}
\SetCommentSty{algocmtsty}
\SetKw{break}{break}
\SetKw{not}{not}
\SetKwFunction{preprocess}{\textsc{Preprocess}}
\SetKwFunction{schedule}{\textsc{Schedule}}
\SetKwFunction{inputGen}{\textsc{InputGen}}
\SetKwFunction{inputEval}{\textsc{InputEval}}
\SetKwFunction{confUpdate}{\textsc{ConfUpdate}}
\SetKwFunction{continue}{\textsc{Continue}}
\SetKwFunction{isBug}{isBug}
\SetKwFunction{getProgram}{getProgram}
\SetKwData{newbugs}{$\bugs^\prime$}
\KwIn{\confs, \timeout}
\KwOut{\bugs \tcp{a finite set of bugs}}
$\bugs \gets \varnothing$\;
$\confs \gets \preprocess{$\confs$}$\;
\While {$\currtime < \timeout \land \continue{\confs}$}{
  \conf $\gets \schedule{\confs, \currtime, \timeout}$\;
  \testcases $\gets \inputGen{\conf}$\;
  \tcp{\bugoracle is embedded in a fuzzer}
  \newbugs, \execinfos $\gets$ \inputEval{\conf, \testcases, \bugoracle}\;
  $\confs \gets \confUpdate{\confs, \conf, \execinfos}$\;
  $\bugs \gets \bugs \cup \newbugs$\;
}
\Return{\bugs}\;
\caption{Fuzz Testing}
\label{algo:fuzzer}
\end{algorithm}

\renewcommand*{\preprocess}{\textsc{\texttt{Preprocess}}\xspace}
\renewcommand*{\schedule}{\textsc{\texttt{Schedule}}\xspace}
\renewcommand*{\inputGen}{\textsc{\texttt{InputGen}}\xspace}
\renewcommand*{\inputEval}{\textsc{\texttt{InputEval}}\xspace}
\renewcommand*{\confUpdate}{\textsc{\texttt{ConfUpdate}}\xspace}
\renewcommand*{\continue}{\textsc{\texttt{Continue}}\xspace}

We present a generic algorithm for fuzz testing, Algorithm~\ref{algo:fuzzer},
which we imagine to have been implemented in a \emph{model fuzzer}.
It is general enough to accommodate existing fuzzing techniques, including
black-, grey-, and white-box fuzzing as defined in~\S\ref{ssec:fuzztax}.
Algorithm~\ref{algo:fuzzer} takes a set of fuzz configurations \confs and a
timeout \timeout as input, and outputs a set of discovered bugs \bugs.
It consists of two parts.
The first part is the \preprocess function, which is executed at the beginning
of a fuzz campaign.
The second part is a series of five functions inside a loop: \schedule,
\inputGen, \inputEval, \confUpdate, and \continue.
Each execution of this loop is called a \emph{fuzz iteration} and each
time \inputEval executes the PUT on a test case is called a \emph{fuzz
  run}.
Note that some fuzzers do \emph{not} implement all five functions.
For example, to model Radamsa~\cite{radamsa}, which never updates the
set of fuzz configurations, \confUpdate always returns the current set
of configurations unchanged.

\begin{description}
  \item[\preprocess\normalfont(\confs)]
    $\rightarrow \confs$ \\
    A user supplies \preprocess with a set of fuzz configurations as input, and
    it returns a potentially-modified set of fuzz configurations.
    Depending on the fuzz algorithm, \preprocess may perform a variety of
    actions such as inserting instrumentation code to PUTs, or measuring the
    execution speed of seed files.  See~\S\ref{sec:preprocess}.

  \item[\schedule\normalfont(\confs, \currtime, \timeout)]
    $\rightarrow \conf$ \\
    \schedule takes in the current set of fuzz configurations, the current time
    \currtime, and a timeout \timeout as input, and selects a fuzz configuration
    to be used for the current fuzz iteration.
    See~\S\ref{sec:schedule}.

  \item[\inputGen\normalfont(\conf)]
    $\rightarrow \testcases$ \\
    \inputGen takes a fuzz configuration as input and returns a set of concrete
    test cases \testcases as output.
    When generating test cases, \inputGen uses specific parameter(s) in \conf.
    Some fuzzers use a seed in \conf for generating test cases, while others use
    a model or grammar as a parameter.
    See~\S\ref{sec:inputgen}.

  \item[\inputEval\normalfont(\conf, \testcases, \bugoracle)]
    $\rightarrow \bugs^\prime, \execinfos$ \\
    \inputEval takes a fuzz configuration \conf, a set of test cases \testcases,
    and a bug oracle \bugoracle as input.
    It executes the PUT on \testcases and checks if the executions violate the
    security policy using the bug oracle \bugoracle.
    It then outputs the set of bugs found $\bugs^\prime$ and
    information about each of the fuzz runs \execinfos, which may be
    used to update the fuzz configurations.
    We assume \bugoracle is embedded in our model fuzzer.
    See~\S\ref{sec:inputeval}.

  \item[\confUpdate\normalfont(\confs, \conf, \execinfos)]
    $\rightarrow \confs$ \\
    \confUpdate takes a set of fuzz configurations \confs, the current
    configuration \conf, and the information about each of the fuzz
    runs \execinfos, as input.
    It may update the set of fuzz configurations \confs. For example, many
    grey-box fuzzers reduce the number of fuzz configurations in \confs based on
    \execinfos.
    See~\S\ref{sec:confupdate}.

  \item[\continue\normalfont(\confs)]
    $\rightarrow \{\texttt{True}, \texttt{False}\}$ \\
    \continue takes a set of fuzz configurations \confs as input and
    outputs a boolean indicating whether a new fuzz iteration should
    occur.
    This function is useful to model white-box fuzzers that can terminate
    when there are no more paths to discover.

\end{description}

\subsection{Taxonomy of Fuzzers}\label{ssec:fuzztax}

For this paper, we have categorized fuzzers into three groups based on
the granularity of semantics a fuzzer observes in each fuzz run.
These three groups are called black-, grey-, and white-box fuzzers,
which we define below.
Note that this classification is different from traditional software
testing, where there are only two major categories (black- and
white-box testing)~\cite{myers:2011}.
As we will discuss in \S\ref{sssec:greybox}, grey-box fuzzing is a variant of
white-box fuzzing that can only obtain some partial information from each fuzz
run.

\subsubsection{\textbf{Black-box Fuzzer}}

The term ``black-box'' is commonly used in software
testing~\cite{myers:2011,beizer:1995} and fuzzing to denote techniques that do
\emph{not} see the internals of the PUT---these techniques can observe only the
input/output behavior of the PUT, treating it as a black-box.
In software testing, black-box testing is also called IO-driven or data-driven
testing~\cite{myers:2011}.
Most traditional
fuzzers~\cite{aitel:blackhat:2001,zzuf,bff-deprecated-merged-with-certfuzz,gpf,foe} are in this category.
Some modern fuzzers, e.g., funfuzz~\cite{funfuzz} and Peach~\cite{peachfuzz},
also take the structural information about inputs into account to generate more
meaningful test cases while maintaining the characteristic of not inspecting the
PUT.
A similar intuition is used in adaptive random testing~\cite{chen:2010}.

\subsubsection{\textbf{White-box Fuzzer}}

At the other extreme of the spectrum, white-box
fuzzing~\cite{godefroid:ndss:2008} generates test cases by analyzing the
internals of the PUT and the information gathered when executing the PUT.
Thus, white-box fuzzers are able to explore the state space of the PUT
systematically.
The term \emph{white-box fuzzing} was introduced by
Godefroid~\cite{godefroid:rt:2007} in 2007 and refers to dynamic symbolic
execution (DSE), which is a variant of symbolic execution~\cite{boyer:1975,
  king:cacm:1976, howden:1975}.
In DSE, symbolic and concrete execution operate concurrently, where concrete
program states are used to simplify symbolic constraints, e.g., concretizing
system calls.
DSE is thus often referred to as \emph{concolic testing} (concrete +
symbolic)~\cite{sen:fse:2005, godefroid:pldi:2005}.
In addition, white-box fuzzing has also been used to describe fuzzers that
employ taint analysis~\cite{ganesh:icse:2009}.
The overhead of white-box fuzzing is typically much higher than that of
black-box fuzzing.
This is partly because DSE implementations~\cite{godefroid:ndss:2008,
  cadar:osdi:2008, avgerinos:icse:2014} often employ dynamic instrumentation and
SMT solving~\cite{moura:cacm:2011}.
While DSE is an active research area~\cite{godefroid:ndss:2008,
  godefroid:pldi:2008, bounimova:icse:2013, pham:ase:2016, jayaraman:2009}, many
DSEs are \emph{not} white-box fuzzers because they do not aim to find security
bugs.
As such, this paper does not provide a comprehensive survey on DSEs and we refer
the reader to recent survey papers~\cite{anand:2013, schwartz:oakland:2010} for
more information on DSEs for non-security applications.

\subsubsection{\textbf{Grey-box Fuzzer}} \label{sssec:greybox}

Some fuzzers~\cite{embleton:blackhat:2006, demott:blackhat:2007,
  takanen:2008} take a middle ground approach which is dubbed
\emph{grey-box fuzzing}.
In general, grey-box fuzzers can obtain \emph{some} information internal to the
PUT and/or its executions.
Unlike white-box fuzzers, grey-box fuzzers do not reason about the
full semantics of the PUT; instead, they may perform lightweight
static analysis on the PUT and/or gather dynamic information about its
executions, such as code coverage.
Grey-box fuzzers rely on approximated, imperfect information in order
to gain speed and be able to test more inputs.
Although there usually is a consensus between security experts, the
distinction between black-, grey- and white-box fuzzing is not always
clear. Black-box fuzzers may collect some information about fuzz runs,
and white-box fuzzers often use some approximations. When classifying
the fuzzers in this survey, particularly in Table~\ref{tab:overview},
we used our best judgement.

An early example of grey-box fuzzer is EFS~\cite{demott:blackhat:2007}, which
uses code coverage gathered from each fuzz run to generate test cases with an
evolutionary algorithm.
Randoop~\cite{pacheco:icse:2007} also used a similar approach, though it did not
target security vulnerabilities.
Modern fuzzers such as AFL~\cite{aflfuzz} and VUzzer~\cite{rawat:ndss:2017} are
exemplars in this category.

\subsection{Fuzzer Genealogy and Overview}

\begin{figure*}[htbp]
\input{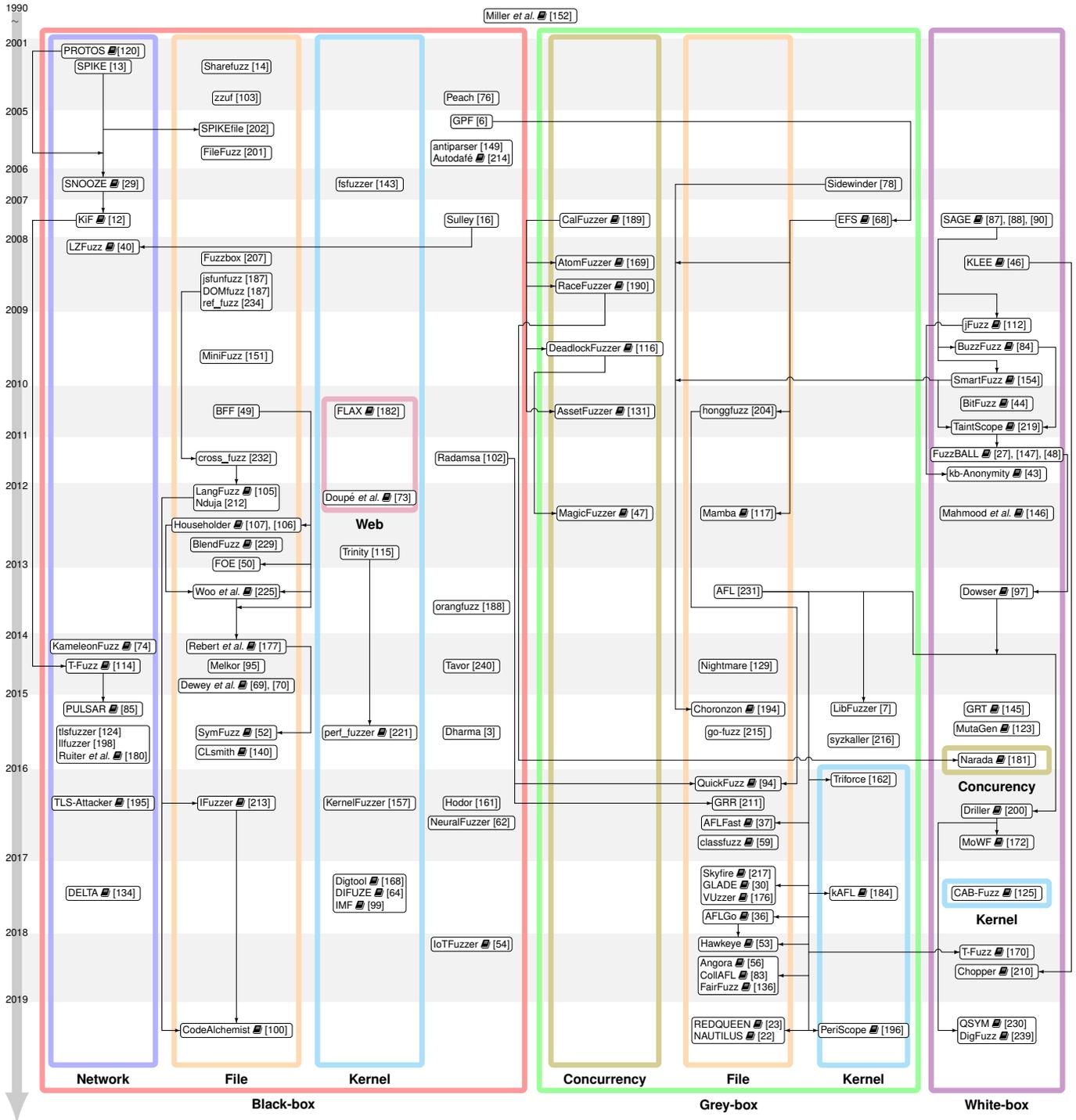}
\caption{Genealogy tracing significant fuzzers' lineage back to
  Miller~\textit{et~al}.'s seminal work. Each node in the same row represents a
  set of fuzzers appeared in the same year. A solid arrow from $X$ to $Y$
  indicates that $Y$ cites, references, or otherwise uses techniques from
  $X$. \faBook{} denotes that a paper describing the work was published.}
\label{fig:genealogy}
\end{figure*}

Figure~\ref{fig:genealogy} (p.~\pageref{fig:genealogy}) presents our
categorization of existing fuzzers in chronological order.
Starting from the seminal work by
Miller~\etal~\cite{miller:cacm:1990}, we manually chose popular
fuzzers that either appeared in a major conference or obtained more
than 100 GitHub stars, and showed their relationships as a graph.
Black-box fuzzers are in the left half of the figure, and grey- and white-box
fuzzers are in the right half.
Furthermore, fuzzers are subdivided depending on the type of input the
PUT uses: file, network, UI, web, kernel I/O, or threads (in the case of
concurrency fuzzers).

\ed{How do these fuzzers relate to the ones in the genealogy?}
Table~\ref{tab:overview} (p.~\pageref{tab:overview}) presents a
detailed summary of the techniques used in the most notable fuzzers in
Figure~\ref{fig:genealogy}. We had to omit some of fuzzers in
Figure~\ref{fig:genealogy} due to space constraints.
Each fuzzer is summarized based on its implementation of the five
functions of our model fuzzer, and a miscellaneous section that
provides other details on the fuzzer.
We describe the properties described by each column below.
The first column (feedback gathering granularity) indicates whether the
fuzzer is black- (\bbfuzz), white- (\wbfuzz), or grey-box (\gbfuzz).
Two circles appear when a fuzzer has two phases which use different
kinds of feedback gathering. For example, SymFuzz~\cite{cha:oakland:2015}
runs a white-box analysis as a preprocessing step in order to optimize
the performance of a subsequent black-box campaign (\bbfuzz+\wbfuzz),
and hybrid fuzzers, such as Driller~\cite{stephens:ndss:2016},
alternate between white- and grey-box fuzzing (\gbfuzz+\wbfuzz).
The second column shows whether the source code of the fuzzer is publicly available.
The third column denotes whether fuzzers need the source code of the PUT to operate.
The fourth column points out whether fuzzers support in-memory fuzzing (see
\S\ref{sssec:inmemory}).
The fifth column is about whether fuzzers can infer models (see
\S\ref{sssec:infmodel}).
The sixth column shows whether fuzzers perform either static or dynamic analysis
in \preprocess.
The seventh column indicates if fuzzers support handling multiple seeds, and
perform scheduling.
The mutation column specifies if fuzzers perform input mutation to generate test
cases. We use \gbfuzz~ to indicate fuzzers that guide input mutation based on the
execution feedback.
The model-based column is about whether fuzzers generate test cases based on a
model.
The constraint-based column shows that fuzzers perform a symbolic analysis to
generate test cases.
The taint analysis column means that fuzzers leverage taint analysis to guide
their test case generation process.
The two columns in the \inputEval section show whether fuzzers perform crash triage
using either stack hashing or code coverage.
The first column of the \confUpdate section indicates if fuzzers
evolve the seed pool during \confUpdate, such as adding new seeds to
the pool (see \S\ref{ssec:evolutionary}).
The second column of the \confUpdate section is about whether fuzzers
learn an input model in an online fashion.
Finally, the third column of the \confUpdate section shows which
fuzzers remove seeds from the seed pool (see \S\ref{ssec:culling}).

\begin{table*}[htbp]
  \centering
  \scriptsize
  \caption{Overview of fuzzers sorted by their instrumentation
  granularity and their name. \bbfuzz, \gbfuzz, and \wbfuzz~ represent
  black-, grey-, and white-box, respectively.}
  \begin{threeparttable}
    \begin{tabularx}{\linewidth}{X ccc |ccc |c |cccc |cc |ccc}
      \toprule

      & \multicolumn{3}{c}{Misc.}
      & \multicolumn{3}{c}{\textbf{\preprocess}}
      & \multicolumn{1}{c}{\textbf{\schedule}}
      & \multicolumn{4}{c}{\textbf{\inputGen}}
      & \multicolumn{2}{c}{\textbf{\inputEval}}
      & \multicolumn{3}{c}{\textbf{\confUpdate}} \\[6pt]

      \textbf{Fuzzer}
      & \multicolumn{1}{c}{\rotatebox[origin=l]{90}{Feedback Gathering Granularity}}
      & \rotatebox[origin=l]{90}{Open-Sourced}
      & \rotatebox[origin=l]{90}{Source Code Required}
      & \multicolumn{1}{c}{\rotatebox[origin=l]{90}{Support In-memory Fuzzing}}
      & \rotatebox[origin=l]{90}{Model Construction}
      & \rotatebox[origin=l]{90}{Program Analysis}
      & \multicolumn{1}{c}{\rotatebox[origin=l]{90}{Seed Scheduling}}
      & \multicolumn{1}{|c}{\rotatebox[origin=l]{90}{Mutation}}
      & \rotatebox[origin=l]{90}{Model-based}
      & \rotatebox[origin=l]{90}{Constraint-based}
      & \rotatebox[origin=l]{90}{Taint Analysis}
      & \multicolumn{1}{c}{\rotatebox[origin=l]{90}{Crash Triage: Stack Hash}}
      & \rotatebox[origin=l]{90}{Crash Triage: Coverage}
      & \multicolumn{1}{c}{\rotatebox[origin=l]{90}{Evolutionary Seed Pool Update}}
      & \rotatebox[origin=l]{90}{Model Update}
      & \rotatebox[origin=l]{90}{Seed Pool Culling} \\


      \midrule

        \rowcolor{lightgray}
        BFF~\cite{bff-deprecated-merged-with-certfuzz}   
      & \bbfuzz   
      & \cm   
      &    
      &    
      &    
      &    
      & \cm 
      & \bbfuzz   
      &    
      &    
      &    
      & \cm   
      &    
      &    
      &    
      & \\[0pt] 

        CodeAlchemist~\cite{han:ndss:2019}   
      & \bbfuzz   
      & \cm   
      &    
      &    
      & \gbfuzz   
      &    
      &    
      &    
      & \cm   
      &    
      &    
      &    
      &    
      &    
      &    
      & \\[0pt] 

        \rowcolor{lightgray}
        CLsmith~\cite{lidbury:pldi:2015}   
      & \bbfuzz   
      & \cm   
      &    
      &    
      &    
      &    
      &   
      & \bbfuzz   
      & \cm  
      &    
      &    
      &    
      &    
      &    
      &    
      & \\[0pt] 

        DELTA~\cite{lee:ndss:2017}   
      & \bbfuzz   
      &    
      &    
      &    
      &    
      &    
      &   
      & \bbfuzz   
      & \cm 
      &    
      &    
      &    
      &    
      &    
      &    
      & \\[0pt] 

        \rowcolor{lightgray}
        DIFUZE~\cite{corina:ccs:2017}   
      & \bbfuzz   
      & \cm   
      & \cm 
      &    
      & \wbfuzz   
      &    
      &    
      & \bbfuzz   
      & \cm   
      &    
      &    
      &    
      &    
      &    
      &    
      & \\[0pt] 

        Digtool~\cite{pan:usenixsec:2017}   
      & \bbfuzz   
      &    
      &    
      &    
      &    
      &    
      &   
      & \bbfuzz   
      &    
      &    
      &    
      &    
      &    
      &    
      &    
      & \\[0pt] 

        \rowcolor{lightgray}
        Doup{\'e}~\etal~\cite{doupe:usec:2012}    
       & \bbfuzz   
       &    
       &    
       &    
       &    
       &    
       &    
       &    
       & \cm   
       &    
       &    
       &    
       &    
       &    
       & \bbfuzz   
       & \\[0pt] 

        FOE~\cite{foe}   
      & \bbfuzz 
      & \cm   
      &    
      &    
      &    
      &    
      & \cm  
      & \bbfuzz   
      &    
      &    
      &    
      & \cm  
      &    
      &    
      &    
      & \\ 

        \rowcolor{lightgray}
        GLADE~\cite{bastani:pldi:2017}   
      & \bbfuzz   
      & \cm   
      &    
      &    
      & \bbfuzz   
      &    
      & \cm   
      &    
      & \cm   
      &    
      &    
      &    
      &    
      &    
      & \bbfuzz   
      & \\[0pt] 

        IMF~\cite{han:ccs:2017}   
      & \bbfuzz   
      & \cm   
      &    
      &    
      & \bbfuzz   
      &    
      &    
      & \bbfuzz   
      & \cm   
      &    
      &    
      &    
      &    
      &    
      &    
      & \\[0pt] 

        \rowcolor{lightgray}
        jsfunfuzz~\cite{funfuzz}   
      & \bbfuzz  
      & \cm   
      &    
      &    
      &    
      &    
      &    
      &    
      & \cm   
      &    
      &    
      & \cm   
      &    
      &    
      &    
      & \\[0pt] 

        LangFuzz~\cite{holler:usec:2012}   
      & \bbfuzz   
      &    
      &    
      &    
      &    
      &    
      &    
      & \bbfuzz   
      & \cm   
      &    
      &    
      &    
      &    
      &    
      &    
      & \\[0pt] 

        \rowcolor{lightgray}
        Miller~\etal~\cite{miller:cacm:1990}   
      & \bbfuzz   
      & \cm   
      &    
      &    
      &    
      &    
      &    
      &    
      &    
      &    
      &    
      &    
      &    
      &    
      &    
      & \\[0pt] 

        Peach~\cite{peachfuzz}   
      & \bbfuzz   
      & \cm   
      &    
      &    
      &    
      &    
      &    
      &    
      & \cm   
      &    
      &    
      & \cm   
      &    
      &    
      &    
      & \\[0pt] 

        \rowcolor{lightgray}
        PULSAR~\cite{gascon:seccomm:2015}   
      & \bbfuzz   
      & \cm   
      &   
      &    
      & \bbfuzz   
      &    
      &    
      &    
      & \cm   
      &    
      &    
      &    
      &    
      &    
      & \bbfuzz   
      & \\[0pt] 

        Radamsa~\cite{radamsa}   
      & \bbfuzz   
      & \cm   
      &    
      &    
      &    
      &    
      &    
      & \bbfuzz   
      & \cm   
      &    
      &    
      &    
      &    
      &    
      &    
      & \\[0pt] 

        \rowcolor{lightgray}
        Ruiter~\etal~\cite{ruiter:usec:2015}   
      & \bbfuzz   
      &    
      &    
      &    
      &    
      &    
      &    
      &    
      & \cm   
      &    
      &    
      &    
      &    
      &    
      & \bbfuzz   
      & \\ 

        TLS-Attacker~\cite{somorovsky:ccs:2016}   
      & \bbfuzz   
      & \cm   
      &    
      &    
      &    
      &    
      &    
      & \bbfuzz   
      &    
      &    
      &    

      &    
      &    
      &    
      &    
      & \\[0pt] 

        \rowcolor{lightgray}
        zuff~\cite{zzuf}   
      & \bbfuzz   
      & \cm   
      &    
      &    
      &    
      &    
      &    
      & \bbfuzz   
      &    
      &    
      &    
      &    
      &    
      &    
      &    
      & \\[0pt] 

        FLAX~\cite{saxena:ndss:2010}   
      & \bbfuzz+\wbfuzz   
      &    
      & \cm 
      &    
      &    
      & \cm 
      &    
      & \bbfuzz   
      &    
      &    
      & \cm 
      &    
      &    
      &    
      &    
      & \\[0pt] 

        \rowcolor{lightgray}
        IoTFuzzer~\cite{chen:ndss:2018}   
      & \bbfuzz+\wbfuzz   
      &    
      &    
      &    
      & \bbfuzz   
      & \cm   
      &    
      & \bbfuzz   
      & \cm   
      &    
      &    
      &    
      &    
      &    
      &    
      & \\ 

        SymFuzz~\cite{cha:oakland:2015}   
      & \bbfuzz+\wbfuzz   
      & \cm   
      &    
      &    
      &    
      & \cm   
      &    
      & \bbfuzz   
      &    
      &    
      &    
      & \cm   
      &    
      &    
      &    
      & \\[0pt] 

        \rowcolor{lightgray}
        AFL~\cite{aflfuzz}   
      & \gbfuzz   
      & \cm   
      &    
      & \cm   
      &    
      &    
      & \cm   
      & \bbfuzz   
      &    
      &    
      &    
      &    
      & \cm   
      & \cm   
      &    
      & \cm \\[0pt] 

        AFLFast~\cite{bohme:ccs:2016}   
      & \gbfuzz   
      & \cm   
      &    
      & \cm   
      &    
      &    
      & \phantom{\tnote{$\dagger$}}\cm\tnote{$\dagger$}   
      & \bbfuzz   
      &    
      &    
      &    
      &    
      & \cm   
      & \cm   
      &    
      & \cm \\[0pt] 

        \rowcolor{lightgray}
        AFLGo~\cite{bohme:ccs:2017}   
      & \gbfuzz   
      & \cm   
      & \cm   
      & \cm   
      &    
      & \cm   
      & \phantom{\tnote{$\dagger$}}\cm\tnote{$\dagger$}   
      & \bbfuzz   
      &    
      &    
      &    
      &    
      & \cm   
      & \cm   
      &    
      & \cm \\[0pt] 

        AssetFuzzer~\cite{lai:icse:2010}   
      & \gbfuzz   
      &     
      & \cm 
      &    
      &    
      & \cm 
      &    
      &    
      &    
      &    
      &    
      &    
      &    
      &    
      &    
      & \\[0pt] 

        \rowcolor{lightgray}
        AtomFuzzer~\cite{park:fse:2008}   
      & \gbfuzz   
      & \cm 
      & \cm 
      &    
      &    
      & \cm 
      &    
      &    
      &    
      &    
      &    
      &    
      &    
      &    
      &    
      & \\[0pt] 

        CalFuzzer~\cite{sen:ase:2007}   
      & \gbfuzz   
      & \cm 
      & \cm 
      &    
      &    
      & \cm 
      &    
      &    
      &    
      &    
      &    
      &    
      &    
      &    
      &    
      & \\[0pt] 

        \rowcolor{lightgray}
      classfuzz~\cite{chen:pldi:2016}   
      & \gbfuzz  
      &    
      &    
      &    
      &    
      &    
      & \cm  
      & \bbfuzz  
      &    
      &    
      &    
      &    
      &    
      &    
      &    
      &  \\[0pt] 

      CollAFL~\cite{gan:oakland:2018}   
      & \phantom{\tnote{$\dagger$}}\gbfuzz\tnote{$\dagger$}   
      &    
      & \cm   
      & \cm   
      &    
      &    
      & \phantom{\tnote{$\dagger$}}\cm\tnote{$\dagger$}   
      & \bbfuzz   
      &    
      &    
      &    
      &    
      & \cm   
      & \cm   
      &    
      & \cm \\[0pt] 

        \rowcolor{lightgray}
        DeadlockFuzzer~\cite{joshi:pldi:2009}   
      & \gbfuzz   
      & \cm 
      & \cm 
      &    
      &    
      & \cm 
      &    
      &    
      &    
      &    
      &    
      &    
      &    
      &    
      &    
      & \\[0pt] 

        FairFuzz~\cite{lemieux:ase:2018}   
      & \gbfuzz   
      & \cm   
      &    
      & \cm   
      &    
      &    
      & \phantom{\tnote{$\dagger$}}\cm\tnote{$\dagger$}   
      & \phantom{\tnote{$\dagger$}}\gbfuzz\tnote{$\dagger$}   
      &    
      &    
      &    
      &    
      & \cm   
      & \cm   
      &    
      & \cm \\[0pt] 

        \rowcolor{lightgray}
        go-fuzz~\cite{gofuzz}   
      & \gbfuzz   
      & \cm   
      & \cm   
      &    
      &    
      &    
      & \cm   
      & \bbfuzz   
      & \cm   
      &    
      &    
      &    
      & \cm   
      & \cm   
      & \gbfuzz   
      & \cm \\[0pt] 

        Hawkeye~\cite{chen:ccs:2018}   
      & \gbfuzz   
      &    
      & \cm   
      & \cm 
      &    
      & \cm   
      & \cm   
      & \gbfuzz  
      &    
      &    
      &    
      &    
      & \cm   
      & \cm   
      &    
      &  \\[0pt] 

        \rowcolor{lightgray}
        honggfuzz~\cite{honggfuzz}   
      &  \gbfuzz  
      &  \cm  
      &    
      &    
      &    
      &    
      &    
      &  \bbfuzz  
      &    
      &    
      &    
      &  \cm  
      &    
      &  \cm  
      &    
      & \\[0pt] 

        kAFL~\cite{schumilo:usenixsec:2017}   
      & \gbfuzz   
      & \cm   
      &    
      &    
      &    
      &    
      &    
      & \bbfuzz   
      &    
      &    
      &    
      &    
      &    
      & \cm   
      &    
      &  \\[0pt] 

        \rowcolor{lightgray}
        LibFuzzer~\cite{libfuzzer}   
      & \gbfuzz   
      & \cm   
      & \cm   
      & \cm   
      &    
      &    
      & \cm   
      & \bbfuzz  
      &    
      &    
      &    
      & 
      & \cm   
      & \cm   
      &    
      & \\[0pt] 

        MagicFuzzer~\cite{cai:icse:2012}   
      & \gbfuzz   
      & \cm 
      & \cm 
      &    
      &    
      & \cm 
      &    
      &    
      &    
      &    
      &    
      &    
      &    
      &    
      &    
      & \\[0pt] 

        \rowcolor{lightgray}
        Nautilus~\cite{aschermann:nautilus:2019}   
      & \gbfuzz   
      & \cm 
      & \cm 
      &    
      &    
      &  
      & \cm   
      & \cm   
      & \cm   
      &    
      &    
      &    
      &    
      & \cm   
      &    
      & \\[0pt] 

        RaceFuzzer~\cite{sen:pldi:2008}   
      & \gbfuzz   
      & \cm 
      & \cm 
      &    
      &    
      & \cm 
      &    
      &    
      &    
      &    
      &    
      &    
      &    
      &    
      &    
      & \\[0pt] 

        \rowcolor{lightgray}
        RedQueen~\cite{aschermann:redqueen:2019}   
      & \gbfuzz  
      & \cm   
      &   
      &    
      &    
      &    
      & \cm   
      & \gbfuzz   
      &    
      &    
      &    
      &    
      &    
      & \cm   
      &    
      & \\[0pt] 

        Steelix~\cite{li:fse:2017}   
      & \phantom{\tnote{$\dagger$}}\gbfuzz\tnote{$\dagger$}   
      &    
      &   
      & \cm  
      &   
      & \cm  
      & \phantom{\tnote{$\dagger$}}\cm\tnote{$\dagger$}  
      & \gbfuzz  
      &   
      &   
      &   
      &   
      & \cm  
      & \phantom{\tnote{$\dagger$}}\cm\tnote{$\dagger$}  
      &   
      & \cm \\[0pt] 

        \rowcolor{lightgray}
        Syzkaller~\cite{syzkaller}   
      & \gbfuzz   
      & \cm   
      & \cm   
      &    
      &    
      &    
      & \cm   
      & \bbfuzz   
      & \cm   
      &    
      &    
      &    
      & \cm   
      & \cm   
      &    
      & \cm \\[0pt] 

        Angora~\cite{chen:oakland:2018}   
      & \gbfuzz+\wbfuzz  
      & \cm   
      & \cm  
      &    
      &    
      &    
      &    
      & \gbfuzz   
      &    
      &    
      & \cm   
      &    
      &    
      & \cm   
      &    
      & \\[0pt] 

        \rowcolor{lightgray}
        Cyberdyne~\cite{goodman:oakland:2018}   
      & \gbfuzz+\wbfuzz   
      & \cm   
      &    
      & \cm   
      &    
      &    
      & \cm   
      & \bbfuzz   
      &    
      & \cm   
      &    
      &    
      & \cm   
      & \cm   
      &    
      & \cm \\[0pt] 

        DigFuzz~\cite{zhao:ndss:2019}   
      & \gbfuzz+\wbfuzz   
      &    
      &    
      &    
      &    
      &    
      & \cm   
      & \bbfuzz   
      &    
      & \cm   
      &    
      &    
      & \cm   
      & \cm   
      &    
      & \cm \\[0pt] 

        \rowcolor{lightgray}
        Driller~\cite{stephens:ndss:2016}   
      & \gbfuzz+\wbfuzz   
      & \cm   
      &    
      &    
      &    
      &    
      & \cm   
      & \bbfuzz   
      &    
      & \cm   
      &    
      &    
      & \cm   
      & \cm   
      &    
      & \cm \\[0pt] 

        QSYM~\cite{yun:usec:2018}   
      & \gbfuzz+\wbfuzz   
      & \cm   
      &    
      &    
      &    
      &    
      & \cm   
      & \bbfuzz   
      &    
      & \cm   
      &    
      &    
      & \cm   
      & \cm   
      &    
      & \cm \\[0pt] 

        \rowcolor{lightgray}
        T-Fuzz~\cite{peng:oakland:2018} 
      & \gbfuzz+\wbfuzz   
      & \cm   
      &    
      & \cm   
      &    
      & \cm   
      & \phantom{\tnote{$\dagger$}}\cm\tnote{$\dagger$}   
      & \bbfuzz   
      &    
      & \cm   
      &    
      &    
      & \cm  
      & \cm   
      &    
      & \cm \\[0pt] 

        VUzzer~\cite{rawat:ndss:2017}   
      & \gbfuzz+\wbfuzz   
      & \cm   
      &    
      &    
      &    
      & \cm   
      & \cm   
      &    
      &    
      &    
      & \cm   
      &    
      &    
      & \cm   
      & \gbfuzz   
      & \\[0pt] 

        \rowcolor{lightgray}
        BitFuzz~\cite{caballero:ccs:2010}   
      & \wbfuzz   
      &    
      &    
      &    
      &    
      & \cm   
      &    
      &    
      &    
      & \cm   
      &    
      &    
      &    
      &    
      &    
      & \\[0pt] 

        BuzzFuzz~\cite{ganesh:icse:2009}   
      & \wbfuzz   
      &    
      & \cm 
      &    
      &    
      & \cm   
      &    
      & \gbfuzz   
      &    
      & \cm   
      & \cm 
      &    
      &    
      &    
      &    
      & \\[0pt] 

        \rowcolor{lightgray}
        CAB-Fuzz~\cite{kim:atc:2017}   
      & \wbfuzz   
      &    
      &    
      &    
      &    
      & \cm   
      & \cm   
      &    
      &    
      & \cm   
      &    
      &    
      &    
      &    
      &    
      & \\[0pt] 

        Chopper~\cite{trabish:icse:2018}   
      & \wbfuzz   
      & \cm   
      & \cm 
      &    
      &    
      & \cm   
      &    
      &    
      &    
      & \cm   
      &    
      &    
      &    
      &    
      &    
      & \\[0pt] 

        \rowcolor{lightgray}
        Dewey~\etal~\cite{dewey:ase:2015}   
      & \wbfuzz   
      & \cm   
      & \cm 
      &    
      &    
      &    
      &    
      &    
      & \cm   
      & \cm   
      &    
      &    
      &    
      &    
      &    
      & \\[0pt] 

        Dowser~\cite{haller:usec:2013}  
      & \wbfuzz   
      &    
      &    
      &    
      &    
      & \cm   
      &    
      &    
      &    
      & \cm   
      & \cm   
      &    
      &    
      &    
      &    
      & \\[0pt] 

        \rowcolor{lightgray}
        GRT~\cite{ma:ase:2015}   
      & \wbfuzz   
      &    
      & \cm 
      &    
      &    
      & \cm   
      & \cm   
      &    
      & \cm   
      &    
      & \cm   
      &    
      &    
      &    
      & \wbfuzz   
      & \\[0pt] 

        KLEE~\cite{cadar:osdi:2008}   
      & \wbfuzz   
      & \cm   
      & \cm 
      &    
      &    
      &    
      &    
      &    
      &    
      & \cm   
      &    
      &    
      &    
      &    
      &    
      & \\[0pt] 

        \rowcolor{lightgray}
        MoWF~\cite{pham:ase:2016}   
      & \wbfuzz   
      &    
      &    
      &    
      &    
      &    
      &    
      &    
      & \cm   
      & \cm   
      &    
      &    
      &    
      &    
      &    
      & \\[0pt] 

        MutaGen~\cite{kargen:fse:2015}   
      & \wbfuzz   
      &    
      &    
      &    
      & \bbfuzz   
      &    
      &    
      & \bbfuzz   
      &    
      &    
      &    
      &    
      &    
      &    
      &    
      & \\[0pt] 

        \rowcolor{lightgray}
        Narada~\cite{samak:pldi:2015}   
      & \wbfuzz   
      & \cm 
      & \cm 
      &    
      &    
      & \cm 
      &    
      &    
      &    
      &    
      &    
      &    
      &    
      &    
      &    
      & \\[0pt] 

        SAGE~\cite{godefroid:ndss:2008}   
      & \wbfuzz   
      &    
      &    
      &    
      &    
      &    
      &    
      &    
      &    
      & \cm  
      &    
      &    
      &    
      &    
      &    
      & \\[0pt] 

        \rowcolor{lightgray}
        TaintScope~\cite{wang:oakland:2010}   
      & \wbfuzz   
      &    
      &    
      &    
      &    
      & \cm   
      & \cm   
      & \gbfuzz   
      &    
      &    
      & \cm   
      &    
      & \cm   
      &    
      &    
      & \\[0pt] 


      \bottomrule
    \end{tabularx}
  \begin{tablenotes}
  \item[$\dagger$] The corresponding fuzzer is derived from AFL, and it changed
    this part of the fuzzing algorithm.
  \end{tablenotes}
  \end{threeparttable}
  \label{tab:overview}
\end{table*}


%
%
%

%
%
%
%
%
%
%
%
%
%
%
%
%
%
%
%
%
%
%
%
%
%
%
%
%
%
%
%
%
%
%
%
%
%
%
%
%
%
%
%
%
%
%
%
%
%
%
%
%
%
%
%
%
%
%
%
%
%
%
%
%
%
%
%

%
%
%

%
%
%
%
%
%
%
%
%
%
%
%
%
%
%
%
%
%
%
%
%
%
%
%
%
%
%
%

%

  %
  %
  %
  %
  %
  %
  %
  %
  %
  %
  %

  %
  %
  %
  %
  %
  %
  %
  %
  %
  %
  %
  %
  %
  %
  %
  %
  %
  %
  %
  %

  %
  %
  %
  %
  %
  %
  %
  %
  %
  %
  %
  %
  %
  %

  %
  %
  %
  %

%
%
%
  %
  %
%
%
%
%
%
%
  %
%
%
%
%
%
%
%
%
%
%
%
%
%
%

%
%
%
%
%
%

%
%
%
%
%

%
\section{Preprocess}\label{sec:preprocess}

Some fuzzers modify the initial set of fuzz configurations before the first
fuzz iteration.
Such preprocessing is commonly used to instrument the PUT, to weed out
potentially-redundant configurations (i.e., ``seed
selection''~\cite{rebert:usec:2014}), to trim seeds, and to generate
driver applications.
As will be detailed in~\S\ref{sssec:premodel} (p.~\pageref{sssec:premodel}),
\preprocess can also be used to prepare a model for future input generation
(\inputGen).

\subsection{Instrumentation} \label{ssec:instrumentation}

Unlike black-box fuzzers, both grey- and white-box fuzzers can
instrument the PUT to gather execution feedback as \inputEval performs
fuzz runs (see \S\ref{sec:inputeval}), or to fuzz the memory contents
at runtime.
The amount of collected information defines the color of a fuzzer
(i.e., black-, white-, or grey-box).
Although there are other ways of acquiring information about the
internals of the PUT (e.g., processor traces or system call
usage~\cite{honggfuzz, goodman:oakland:2018}), instrumentation is
often the method that collects the most valuable feedback.

Program instrumentation can be either static or dynamic---the former
happens before the PUT runs (\preprocess), whereas the latter happens
while the PUT is running (\inputEval).
But for the reader's convenience, we summarize
both static and dynamic instrumentation in this section.

Static instrumentation is often performed at compile time on either source code
or intermediate code.
Since it occurs before runtime, it generally imposes less runtime
overhead than dynamic instrumentation.
If the PUT relies on libraries, these have to be separately instrumented,
commonly by recompiling them with the same instrumentation.
Beyond source-based instrumentation, researchers have also developed
binary-level static instrumentation (i.e., binary rewriting)
tools~\cite{zhang:vee:2014, laurenzano:pebil:2010, edwards:vulcan:2001}.

Although it has higher overhead than static instrumentation, dynamic
instrumentation has the advantage that it can easily instrument dynamically
linked libraries, because the instrumentation is performed at runtime.
There are several well-known dynamic instrumentation tools such as
DynInst~\cite{dyninst}, DynamoRIO~\cite{bruening:2004},
Pin~\cite{luk:pldi:2005}, Valgrind~\cite{nethercote:pldi:2007}, and
QEMU~\cite{bellard:atc:2005}.

A given fuzzer can support more than one type of instrumentation.
For example, AFL supports static instrumentation at the source code level with a
modified compiler, or dynamic instrumentation at the binary level with the help
of QEMU~\cite{bellard:atc:2005}.
When using dynamic instrumentation, AFL can either instrument (1)
executable code in the PUT itself, which is the default setting, or
(2) executable code in the PUT and any external libraries (with the
\aflinstlibs option).
The second option---instrumenting all encountered code---can report coverage
information for code in external libraries, and thus provides a more complete
picture of coverage.
However, this will also encourage AFL to fuzz additional paths in
external library functions.

\subsubsection{Execution Feedback}

Grey-box fuzzers typically take execution feedback as input to evolve test
cases. AFL and its descendants compute branch coverage by instrumenting every
branch instruction in the PUT. However, they store the branch coverage
information in a bit vector, which can cause path
collisions.
CollAFL~\cite{gan:oakland:2018} recently addressed this issue by introducing a
new path-sensitive hash function.
Meanwhile, LibFuzzer~\cite{libfuzzer} and Syzkaller~\cite{syzkaller} use node
coverage as their execution feedback.
Honggfuzz~\cite{honggfuzz} allows users to choose which execution feedback to
use.

\subsubsection{In-Memory Fuzzing} \label{sssec:inmemory}

When testing a large program, it is sometimes desirable to fuzz
\emph{only a portion} of the PUT without re-spawning a process for
each fuzz iteration in order to minimize execution overhead. For
example, complex (e.g., GUI) applications often require several
seconds of processing before they accept input.  One approach to
fuzzing such programs is to take a snapshot of the PUT after the GUI
is initialized. To fuzz a new test case, one can then restore the
memory snapshot before writing the new test case directly into memory
and executing it. The same intuition applies to fuzzing network
applications that involve heavy interaction between client and server.
This technique is called in-memory
fuzzing~\cite{hoglund:blackhat:2003}.
As an example, GRR~\cite{grr,goodman:oakland:2018} creates a snapshot before
loading any input bytes. This way, it can skip over unnecessary startup code.
AFL also employs a fork server to avoid some of the process startup
costs. Although it has the same motivation as in-memory fuzzing, a
fork server involves forking off a new process for every fuzz
iteration (see \S\ref{sec:inputeval}).

Some fuzzers~\cite{libfuzzer,aflfuzz} perform in-memory fuzzing on a
function without restoring the PUT's state after each iteration. We call such a
technique as an \emph{in-memory API fuzzing}. For example, AFL has an
option called persistent mode~\cite{lcamtuf:persistent}, which
repeatedly performs in-memory API fuzzing in a loop without restarting
the process. In this case, AFL ignores potential side effects from the
function being called multiple times in the same execution.

Although efficient, in-memory API fuzzing suffers from unsound fuzzing
results: bugs (or crashes) found from in-memory fuzzing may \emph{not}
be reproducible, because (1) it is not always feasible to construct a
valid calling context for the target function, and (2) there can be
side-effects that are not captured across multiple function
calls. Notice that the soundness of in-memory API fuzzing mainly
depends on the entry point function, and finding such a function is a
challenging task.

\subsubsection{Thread Scheduling}

Race condition bugs can be difficult to trigger because they rely on
non-deterministic behaviors which may only occur infrequently.
However, instrumentation can also be used to trigger different
non-deterministic program behaviors by explicitly controlling how
threads are
scheduled~\cite{sen:ase:2007,sen:pldi:2008,park:fse:2008,joshi:pldi:2009,lai:icse:2010,cai:icse:2012,samak:pldi:2015}. Existing
work has shown that even randomly scheduling threads can be effective
at finding race condition bugs~\cite{sen:ase:2007}.

\subsection{Seed Selection}\label{ssec:seedselection}

Recall from \S\ref{sec:def} that fuzzers receive a set of fuzz
configurations that control the behavior of the fuzzing algorithm.
Unfortunately, some parameters of fuzz configurations, such as seeds
for mutation-based fuzzers, have large value domains.
For example, suppose an analyst fuzzes an MP3 player that accepts MP3 files as
input. There is an unbounded number of valid MP3 files, which raises a natural
question: which seeds should we use for fuzzing?
This problem of decreasing the size of the initial seed pool is known as the
\emph{seed selection problem}~\cite{rebert:usec:2014}.

There are several approaches and tools that address the seed selection
problem~\cite{rebert:usec:2014,peachfuzz}. A common approach is to find
a minimal set of seeds that maximizes a coverage metric, e.g., node
coverage, and this process is called computing a \emph{minset}.
For example, suppose the current set of configurations \confs consists
of two seeds $s_1$ and $s_2$ that cover the following addresses of the
PUT:
$\left\{ s_1 \rightarrow \left\{ 10, 20 \right\}, s_2 \rightarrow
  \left\{ 20, 30 \right\} \right\}$.
If we have a third seed $s_3 \rightarrow \left\{ 10, 20, 30 \right\}$
that executes roughly as fast as $s_1$ and $s_2$, one could argue it
makes sense to fuzz $s_3$ instead of $s_1$ and $s_2$, since it
intuitively tests more program logic for half the execution time cost.
This intuition is supported by Miller's
report~\cite{miller:cansecwest:2008}, which showed that a 1\% increase
in code coverage increased the percentage of bugs found by .92\%.
As is noted in \S\ref{ssec:culling}, this step can also be part of
\confUpdate, which is useful for fuzzers that can introduce new seeds
into the seed pool throughout the campaign.

Fuzzers use a variety of different coverage metrics in practice. For
example, AFL's minset is based on branch coverage with a logarithmic
counter on each branch. The rationale behind this decision is to allow
branch counts to be considered different only when they differ in
orders of magnitude. %
Honggfuzz~\cite{honggfuzz} computes coverage based on the number of
executed instructions, executed branches, and unique basic
blocks. This metric allows the fuzzer to add longer executions to the
minset, which can help discover denial of service vulnerabilities or
performance problems.

\subsection{Seed Trimming}\label{sec:seed-trimming}

Smaller seeds are likely to consume less memory and entail higher
throughput. Therefore, some fuzzers attempt to reduce the size of
seeds prior to fuzzing them, which is called \emph{seed
  trimming}. Seed trimming can happen prior to the main fuzzing loop
in \preprocess or as part of \confUpdate.  One notable fuzzer that
uses seed trimming is AFL~\cite{aflfuzz}, which uses its code coverage
instrumentation to iteratively remove a portion of the seed as long as
the modified seed achieves the same coverage.
Meanwhile, Rebert~\etal~\cite{rebert:usec:2014} reported that their
size minset algorithm, which selects seeds by giving higher priority
to smaller seeds in size, results in fewer unique bugs
compared to a random seed selection.
For the specific case of fuzzing Linux system call handlers,
MoonShine~\cite{pailoor:usec:2018} extends syzkaller~\cite{syzkaller}
to reduce the size of seeds while preserving the dependencies between
calls which are detected using a static analysis.

\subsection{Preparing a Driver Application}

When it is difficult to directly fuzz the PUT, it makes sense to prepare a
driver for fuzzing. This process is largely manual in practice although this is
done only once at the beginning of a fuzzing campaign.
For example, when our target is a library, we need to prepare for a driver
program that calls functions in the library.
Similarly, kernel fuzzers may fuzz userland applications to test
kernels~\cite{kim:atc:2017, dmytro:ioctlfuzzer, beer:iokitfuzzer}.
IoTFuzzer~\cite{chen:ndss:2018} targets IoT devices by letting the driver communicate with
the corresponding smartphone application.

\section{Scheduling}\label{sec:schedule}

In fuzzing, scheduling means selecting a fuzz configuration for the
next fuzz iteration.
As we have explained in \S\ref{ssec:fuzzdef}, the content of each configuration
depends on the type of the fuzzer.
For simple fuzzers, scheduling can be straightforward---for example,
zzuf~\cite{zzuf} in its default mode allows only one configuration and
thus there is simply no decision to make.
But for more advanced fuzzers such as
BFF~\cite{bff-deprecated-merged-with-certfuzz} and
AFLFast~\cite{bohme:ccs:2016}, a major factor to their success lies in their
innovative scheduling algorithms.
In this section, we will discuss scheduling algorithms for black- and grey-box
fuzzing only; scheduling in white-box fuzzing requires a complex setup unique to
symbolic executors and we refer the reader to another source~\cite{bounimova:icse:2013}.

\subsection{The Fuzz Configuration Scheduling (FCS) Problem}\label{sec:fcs}

The goal of scheduling is to analyze the currently-available information about
the configurations and pick one that is likely to lead to the most favorable
outcome, e.g., finding the most number of unique bugs, or maximizing the
coverage attained by the set of generated inputs.
Fundamentally, every scheduling algorithm confronts the same \emph{exploration
  vs. exploitation} conflict---time can either be spent on gathering more
accurate information on each configuration to inform future decisions (explore),
or on fuzzing the configurations that are currently believed to lead to more
favorable outcomes (exploit).
Woo \etal~\cite{woo:ccs:2013} dubbed this inherent conflict the Fuzz
Configuration Scheduling (FCS) Problem.

In our model fuzzer (Algorithm~\ref{algo:fuzzer}), the function \schedule
selects the next configuration based on (i) the current set of fuzz
configurations $\confs$, (ii) the current time $\currtime$, and (iii) the total
time budget $\timeout$.
This configuration is then used for the next fuzz iteration.
Notice that \schedule is only about decision-making. The information
on which this decision is based is acquired by \preprocess and
\confUpdate, which augment \confs with this knowledge.

\subsection{Black-box FCS Algorithms}\label{sec:fcs-blackbox}

In the black-box setting, the only information an FCS algorithm can use is the
fuzz outcomes of a configuration---the number of crashes and bugs found with it
and the amount of time spent on it so far.
Householder and Foote~\cite{householder:bff:2012} were the first to study how such
information can be leveraged in the CERT BFF black-box mutational
fuzzer~\cite{bff-deprecated-merged-with-certfuzz}.
They postulated that a configuration with a higher observed success rate (\#bugs
/ \#runs) should be preferred.
Indeed, after replacing the uniform-sampling scheduling algorithm in BFF, they
observed 85\percent more unique crashes over 5 million runs of ffmpeg,
demonstrating the potential benefit of more advanced FCS algorithms.

Shortly after, the above idea was improved on multiple
fronts by Woo \etal~\cite{woo:ccs:2013}.
First, they refined the mathematical model of black-box mutational fuzzing from
a sequence of Bernoulli trials~\cite{householder:bff:2012} to the \emph{Weighted
  Coupon Collector's Problem with Unknown Weights} (WCCP/UW).
Whereas the former assumes each configuration has a fixed eventual success
probability and learns it over time, the latter explicitly maintains an
upper-bound on this probability as it decays.
Second, the WCCP/UW model naturally leads Woo \etal to investigate algorithms
for \emph{multi-armed bandit} (MAB) problems, which is a popular formalism to
cope with the exploration vs. exploitation conflict in decision
science~\cite{berry:1985}.
To this end, they were able to design MAB algorithms to accurately exploit
configurations that are not known to have decayed yet.
Third, they observed that, all else being equal, a configuration that is faster
to fuzz allows a fuzzer to either collect more unique bugs with it, or decrease
the upperbound on its future success probability more rapidly.
This inspired them to normalize the success probability of a configuration by
the time that has been spent on it, thus causing a faster configuration to be
more preferable.
Fourth, they changed the orchestration of fuzz runs in BFF from a
fixed number of fuzz iterations per configuration selection (``epochs'' in BFF
parlance) to a fixed amount of time per selection.
With this change, BFF is no longer forced to spend more time in a slow
configuration before it can re-select.
By combining the above, the evaluation~\cite{woo:ccs:2013} showed a
$1.5\times\!$ increase in the number of unique bugs found using the
same amount of time as the existing BFF.

\subsection{Grey-box FCS Algorithms}\label{sec:fcs-greybox}

In the grey-box setting, an FCS algorithm can choose to use a richer set of
information about each configuration, e.g., the coverage attained when fuzzing a
configuration.
AFL~\cite{aflfuzz} is the forerunner in this category and it is based on an
evolutionary algorithm (EA).
Intuitively, an EA maintains a population of configurations, each with some
value of ``fitness''.
An EA selects fit configurations and applies them to genetic transformations
such as mutation and recombination to produce offspring, which may later become
new configurations.
The hypothesis is that these produced configurations are more likely to be fit.

To understand FCS in the context of an EA, we need to define (i) what makes a
configuration fit, (ii) how configurations are selected, and (iii) how a
selected configuration is used.
As a high-level approximation, among the configurations that exercise a
control-flow edge, AFL considers the one that contains the fastest and smallest
input to be fit (``favorite'' in AFL parlance).
AFL maintains a queue of configurations, from which it selects the next fit
configuration \emph{essentially} as if the queue is circular.
Once a configuration is selected, AFL fuzzes it for essentially a constant
number of runs.
From the perspective of FCS, notice that the preference for fast configurations
is common for the black-box setting~\cite{woo:ccs:2013}.

Recently, AFLFast by B\"ohme \etal~\cite{bohme:ccs:2016} has improved upon AFL
in each of the three aspects above.
First, AFLFast adds two overriding criteria for an input to become a
``favorite'':
(i) Among the configurations that exercise a control-flow edge, AFLFast favors
the one that has been chosen least.
This has the effect of cycling among configurations that exercise this edge,
thus increasing exploration.
(ii) When there is a tie in (i), AFLFast favors the one that exercises a path
that has been exercised least.
This has the effect of increasing the exercise of rare paths, which may uncover
more unobserved behavior.
Second, AFLFast forgoes the round-robin selection in AFL and instead selects the
next fit configuration based on a priority.
In particular, a fit configuration has a higher priority than another if it has
been chosen less often or, when tied, if it exercises a path that has been
exercised less often.
In the same spirit as the first change, this has the effect of increasing the
exploration among fit configurations and the exercising of rare paths.
Third, AFLFast fuzzes a selected configuration a variable number of times as
determined by a \emph{power schedule}.
The FAST power schedule in AFLFast starts with a small ``energy'' value to
ensure initial exploration among configurations and increases exponentially up
to a limit to quickly ensure sufficient exploitation.
In addition, it also normalizes the energy by the number of generated inputs
that exercise the same path, thus promoting explorations of less-frequently
fuzzed configurations.
The overall effect of these changes is very significant---in a 24-hour
evaluation, B\"ohme \etal observed AFLFast discovered 3 bugs that AFL did not,
and was on average 7$\times$ faster than AFL on 6 other bugs that were
discovered by both.

AFLGo~\cite{bohme:ccs:2017} extends AFLFast by modifying its priority
attribution in order to target specific program locations.
Hawkeye~\cite{chen:ccs:2018} further improves directed fuzzing by leveraging
a static analysis in its seed scheduling and input generation.
FairFuzz~\cite{lemieux:ase:2018} guides the campaign to exercise rare branches
by employing a mutation mask for each pair of a seed and a rare branch.
QTEP~\cite{wang:fse:2017} uses static analysis to infer which part of the binary
is more `faulty' and prioritize configurations that cover them.

\section{Input Generation}\label{sec:inputgen}

Since the content of a test case directly controls whether or not a
bug is triggered, the technique used for \emph{input generation} is
naturally one of the most influential design decisions in a fuzzer.
Traditionally, fuzzers are categorized into either generation- or
mutation-based fuzzers~\cite{sutton:2007}. Generation-based fuzzers
produce test cases based on a given model that describes the inputs
expected by the PUT. We call such fuzzers {\em model-based} fuzzers in
this paper.
On the other hand, mutation-based fuzzers produce test cases by mutating a given
\emph{seed} input.
Mutation-based fuzzers are generally considered to be {\em model-less}
because seeds are merely example inputs and even in large numbers they
do not completely describe the expected input space of the PUT.
In this section, we explain and classify the various input generation techniques
used by fuzzers based on the underlying test case generation (\inputGen)
mechanism.

\subsection{Model-based (Generation-based) Fuzzers}

Model-based fuzzers generate test cases based on a given model that
describes the inputs or executions that the PUT may accept, such as a
grammar precisely characterizing the input format or less precise
constraints such as magic values identifying file types.

\subsubsection{Predefined Model} \label{sssec:premodel}

Some fuzzers use a model that can be configured by the user. For
example, Peach~\cite{peachfuzz}, PROTOS~\cite{kaksonen:2001}, and
Dharma~\cite{dharma} take in a specification provided by the user.
Autodaf\'{e}~\cite{vuagnoux:ccc:2005}, Sulley~\cite{sulley},
SPIKE~\cite{aitel:blackhat:2001}, and SPIKEfile~\cite{sutton:blackhat:2005} expose
APIs that allow analysts to create their own input
models. Tavor~\cite{tavor} also takes in an
input specification written in Extended Backus-Naur form (EBNF) and
generates test cases conforming to the corresponding grammar.
Similarly, network protocol fuzzers such as
PROTOS~\cite{kaksonen:2001}, SNOOZE~\cite{banks:isc:2006},
KiF~\cite{abdelnur:iptcomm:2007}, and T-Fuzz~\cite{johansson:icst:2014}
also take in a protocol specification from the user.
Kernel API
fuzzers~\cite{trinity,syzkaller,kernelfuzzer,triforce,weaver:2015}
define an input model in the form of system call templates. These
templates commonly specify the number and types of arguments a system
call expects as inputs. The idea of using a model in kernel fuzzing
originated in Koopman~\etal's seminal work~\cite{koopman:srds:1997}
where they compared the robustness of OSes with a finite set of
manually chosen test cases for system calls.
Nautilus~\cite{aschermann:nautilus:2019} employs grammar-based input
generation for general-purpose fuzzing, and also
uses its grammar for seed trimming (see~\S\ref{sec:seed-trimming}).

Other model-based fuzzers target a specific language or grammar, and
the model of this language is built in to the fuzzer itself.
For example, cross\_fuzz~\cite{crossfuzz} and DOMfuzz~\cite{funfuzz}
generate random Document Object Model (DOM) objects.
Likewise, jsfunfuzz~\cite{funfuzz} produces random, but syntactically
correct JavaScript code based on its own grammar model.
QuickFuzz~\cite{grieco:2016} utilizes existing Haskell libraries that
describe file formats when generating test cases.
Some network protocol fuzzers such as Frankencerts~\cite{brubaker:oakland:2014},
TLS-Attacker~\cite{somorovsky:ccs:2016}, tlsfuzzer~\cite{tlsfuzzer}, and
llfuzzer~\cite{llfuzzer} are designed with models of specific network protocols
such as TLS and NFC.
Dewey~\etal~\cite{dewey:ase:2014,dewey:ase:2015} proposed a way
to generate test cases that are not only grammatically correct, but
also semantically diverse by leveraging constraint logic programming.
LangFuzz~\cite{holler:usec:2012} produces code fragments by parsing a
set of seeds that are given as input. It then randomly combines the
fragments, and mutates seeds with the fragments to generate test
cases.
Since it is provided with a grammar, it always produces syntactically
correct code. LangFuzz was applied to JavaScript and PHP.
BlendFuzz~\cite{yang:trustcom:2012} is based on similar ideas as
LangFuzz, but targets XML and regular expression parsers.

\subsubsection{Inferred Model} \label{sssec:infmodel}

Inferring the model rather than relying on a predefined or
user-provided model has recently been gaining traction.
Although there is an abundance of published research on the topic of
automated input format and protocol reverse
engineering~\cite{cui:ccs:2008,caballero:ccs:2007,lin:fse:2008,comparetti:oakland:2009,bai:ndss:2013},
only a few fuzzers leverage these techniques.
Similar to instrumentation (\S\ref{ssec:instrumentation}), model
inference can occur in either \preprocess or \confUpdate.

\paragraph{Model Inference in \preprocess}
Some fuzzers infer the model as a preprocessing step.
TestMiner~\cite{dellatoffola:ase:2017} searches for the data available in the
PUT, such as literals, to predict suitable inputs.
Given a set of seeds and a grammar, Skyfire~\cite{wang:oakland:2017} uses a
data-driven approach to infer a probabilitistic context-sensitive grammar, and
then uses it to generate a new set of seeds.
Unlike previous works, it focuses on generating semantically valid inputs.
IMF~\cite{han:ccs:2017} learns a kernel API model by analyzing system
API logs, and it produces C code that invokes a sequence of API calls
using the inferred model.
CodeAlchemist~\cite{han:ndss:2019} breaks JavaScript code into ``code
bricks'', and computes assembly constraints, which describe when
distinct bricks can be assembled or merged together to produce
semantically valid test cases.  These constraints are computed using
both a static analysis and dynamic analysis.
Neural~\cite{neural} and {L}earn\&{F}uzz~\cite{godefroid:ase:2017} use a neural
network-based machine learning technique to learn a model from a given set of
test files, and generate test cases from the inferred
model. Liu~\etal~\cite{liu:icse:2017} proposed a similar approach specific to
text inputs.

\paragraph{Model Inference in \confUpdate}
Other fuzzers can potentially update their model after each fuzz iteration.
PULSAR~\cite{gascon:seccomm:2015} automatically infers a network protocol model
from a set of captured network packets generated from a program. The learned
network protocol is then used to fuzz the program. PULSAR internally builds a
state machine, and maps which message token is correlated with a state. This
information is later used to generate test cases that cover more states in the
state machine.
Doup{\'e}~\etal~\cite{doupe:usec:2012} propose a way to infer the state
machine of a web service by observing the I/O behavior. The inferred
model is then used to scan for web vulnerabilities.
The work of Ruiter~\etal~\cite{ruiter:usec:2015} is similar, but targets TLS and bases
its implementation on LearnLib~\cite{raffelt:fmics:2005}.
GLADE~\cite{bastani:pldi:2017} synthesizes a context-free
grammar from a set of I/O samples, and fuzzes the PUT using the inferred
grammar.
Finally, go-fuzz~\cite{gofuzz} is a grey-box fuzzer, which builds a model for
each of the seed it adds to its seed pool. This model is used to generate new
inputs from this seed.%

\subsubsection{Encoder Model}

Fuzzing is often used to test \emph{decoder} programs which parse a
certain file format.
Many file formats have corresponding \emph{encoder} programs,
which can be thought of as an implicit model of the file format.
MutaGen~\cite{kargen:fse:2015} is a unique fuzzer that leverages this
implicit model contained in an encoder program to generate new test
cases.
MutaGen leverages mutation to generate test cases, but unlike most
mutation-based fuzzers, which mutate an existing \emph{test case}
(see~\S\ref{sec:mutation-based-fuzzers}), MutaGen mutates the
\emph{encoder program}.
Specifically, to produce a new test case MutaGen computes a dynamic
program slice of the encoder program and runs it.  The underlying idea
is that the program slices will slightly change the behavior of the
encoder program so that it produces testcases that are slightly
malformed.

\subsection{Model-less (Mutation-based) Fuzzers}
\label{sec:mutation-based-fuzzers}

Classic random testing~\cite{arcuri:2012,hamlet:2006} is not efficient
in generating test cases that satisfy specific path
conditions. Suppose there is a simple C statement: \texttt{if (input
== 42)}.
If \texttt{input} is a 32-bit integer, the probability of randomly guessing the
right input value is $1/2^{32}$.
The situation gets worse when we consider well-structured input such as an MP3
file. It is extremely unlikely that random testing will generate a valid MP3
file as a test case in a reasonable amount of time.
As a result, the MP3 player will mostly reject the generated test cases from
random testing at the parsing stage before reaching deeper parts of the program.

This problem motivates the use of seed-based input generation as well
as white-box input generation (see~\S\ref{ssec:whitebox}).
Most model-less fuzzers use a \emph{seed}, which is an input to the PUT, in
order to generate test cases by modifying the seed.
A seed is typically a well-structured input of a type supported by the PUT: a
file, a network packet, or a sequence of UI events.
By mutating only a fraction of a valid file, it is often possible to generate a
new test case that is mostly valid, but also contains abnormal values to trigger
crashes of the PUT. There are a variety of methods used to mutate seeds, and we
describe the common ones below.

\subsubsection{Bit-Flipping}\label{sssec:bitflip}

Bit-flipping is a common technique used by many model-less
fuzzers~\cite{aflfuzz,honggfuzz,zzuf,gpf,radamsa}. Some fuzzers simply
flip a fixed number of bits, while others determine the number of bits
to flip at random. To randomly mutate seeds, some fuzzers employ a
user-configurable parameter called the \emph{mutation ratio}, which
determines the number of bit positions to flip for a single execution
of \inputGen. Suppose a fuzzer wants to flip $K$ random bits from a
given $N$-bit seed. In this case, a mutation ratio of the fuzzer is
$K/N$.

SymFuzz~\cite{cha:oakland:2015} showed that fuzzing performance is sensitive to
the mutation ratio, and that there is not a single ratio that works well for all
PUTs. There are several approaches to find a good mutation ratio.
BFF~\cite{bff-deprecated-merged-with-certfuzz} and FOE~\cite{foe} use an
exponentially scaled set of mutation ratios for each seed and allocate more
iterations to mutation ratios that prove to be statistically
effective~\cite{householder:bff:2012}. SymFuzz~\cite{cha:oakland:2015} leverages
a white-box program analysis to infer a good mutation ratio for each seed.

\subsubsection{Arithmetic Mutation}\label{sssec:arith}

AFL~\cite{aflfuzz} and honggfuzz~\cite{honggfuzz} contain another mutation
operation where they consider a selected byte sequence as an integer, and
perform simple arithmetic on that value.
The computed value is then used to replace the selected byte
sequence.
The key intuition is to bound the effect of mutation by a
small number.
For example, AFL selects a 4-byte value from a seed, and treats the
value as an integer $i$. It then replaces the value in the seed with
$i \pm r$, where $r$ is a randomly generated small integer. The range
of $r$ depends on the fuzzer, and is often user-configurable. In AFL,
the default range is: $0 \le r < 35$.

\subsubsection{Block-based Mutation}\label{sssec:block}

There are several block-based mutation methodologies, where a block is
a sequence of bytes of a seed:
(1) insert a randomly generated block into a random position of a
seed~\cite{aflfuzz,libfuzzer};
(2) delete a randomly selected block from a
seed~\cite{aflfuzz,radamsa,honggfuzz,libfuzzer};
(3) replace a randomly selected block with a random
value~\cite{aflfuzz,honggfuzz,radamsa,libfuzzer};
(4) randomly permute the order of a sequence of
blocks~\cite{radamsa,libfuzzer};
(5) resize a seed by appending a random block~\cite{honggfuzz}; and
(6) take a random block from a seed to insert/replace a random block
of another seed~\cite{aflfuzz,libfuzzer}.

\subsubsection{Dictionary-based Mutation}\label{sssec:dict}

Some fuzzers use a set of predefined values with potentially
significant semantic weight, e.g., 0 or $-1$, and format strings for
mutation. For example, AFL~\cite{aflfuzz}, honggfuzz~\cite{honggfuzz},
and LibFuzzer~\cite{libfuzzer} use values such as 0, -1, and 1 when
mutating integers.  Radamsa~\cite{radamsa} employs Unicode strings and
GPF~\cite{gpf} uses formatting characters such as \texttt{\%x} and
\texttt{\%s} to mutate strings~\cite{formatstring}.

\subsection{White-box Fuzzers}\label{ssec:whitebox}

White-box fuzzers can also be categorized into either model-based or
model-less fuzzers. For example, traditional dynamic symbolic
execution~\cite{godefroid:ndss:2008,jayaraman:2009,babic:issta:2011,martignoni:asplos:2012,stephens:ndss:2016}
does not require any model as in mutation-based fuzzers, but some
symbolic executors~\cite{godefroid:pldi:2008,pham:ase:2016,kim:atc:2017}
leverage input models such as an input grammar to guide the symbolic
executor.

Although many white-box fuzzers including the seminal work by
Godefroid~\etal~\cite{godefroid:ndss:2008} use dynamic symbolic
execution to generate test cases, not all white-box fuzzers are
dynamic symbolic executors.
Some fuzzers~\cite{wang:oakland:2010,cha:oakland:2015,ma:ase:2015,
saxena:ndss:2010} leverage a white-box program analysis to find information
about the inputs a PUT accepts in order to use it with black- or grey-box
fuzzing.
In the rest of this subsection, we briefly summarize the existing white-box
fuzzing techniques based on their underlying test case algorithm.
Please note that we intentionally omit dynamic symbolic executors such
as \cite{godefroid:pldi:2005, sen:fse:2005, chipounov:asplos:2011, cadar:osdi:2008,
  tillmann:tap:2008, cha:oakland:2012} unless they explicitly call
themselves as a fuzzer as mentioned in \S\ref{ssec:paper-select}.

\subsubsection{Dynamic Symbolic Execution}

At a high level, classic symbolic
execution~\cite{king:cacm:1976,boyer:1975,howden:1975} runs a program
with symbolic values as inputs, which represents all possible
values. As it executes the PUT, it builds symbolic expressions instead
of evaluating concrete values. Whenever it reaches a conditional
branch instruction, it conceptually forks two symbolic interpreters,
one for the true branch and another for the false branch. For every
path, a symbolic interpreter builds up a path formula (or path
predicate) for every branch instruction it encountered during an
execution. A path formula is satisfiable if there is a concrete input
that executes the desired path. One can generate concrete inputs by
querying an SMT solver \cite{moura:cacm:2011} for a solution to a path
formula.
Dynamic symbolic execution is a variant of traditional symbolic execution, where
both symbolic execution and concrete execution operate at the same time. Thus,
we often refer to dynamic symbolic execution as concolic (concrete + symbolic)
testing. The idea is that concrete execution states can help reduce the
complexity of symbolic constraints.
An extensive review of the academic literature of dynamic symbolic execution,
beyond its application to fuzzing, is out of the scope of this paper. However, a
broader treatment of dynamic symbolic execution can be found
in other sources~\cite{anand:2013,schwartz:oakland:2010}.

Dynamic symbolic execution is slow compared to grey-box or black-box approaches
as it involves instrumenting and analyzing every single instruction of the PUT.
To cope with the high cost, a common strategy has been to narrow its
usage; for instance, by letting the user to specify uninteresting parts of the
code~\cite{trabish:icse:2018}, or
by alternating between concolic testing and grey-box fuzzing.
Driller~\cite{stephens:ndss:2016} and Cyberdyne~\cite{goodman:oakland:2018} have
shown the usefulness of this technique at the DARPA Cyber Grand Challenge.
QSYM~\cite{yun:usec:2018} seeks to improve the integration between grey- and white-box fuzzing by
implementing a fast concolic execution engine.
DigFuzz~\cite{zhao:ndss:2019} optimizes the switch between grey- and
white-box fuzzing by first estimating the probability of exercising
each path using grey-box fuzzing, and then having its white-box fuzzer
focus on the paths that are believed to be most challenging for
grey-box fuzzing.

\subsubsection{Guided Fuzzing}

Some fuzzers leverage static or dynamic program analysis techniques for enhancing the
effectiveness of fuzzing.
These techniques usually involve fuzzing in two phases: (i) a costly program
analysis for obtaining useful information about the PUT, and (ii) test case
generation with the guidance from the previous analysis.
This is denoted in the sixth column of Table~\ref{tab:overview}
(p.~\pageref{tab:overview}).
For example, TaintScope~\cite{wang:oakland:2010} uses a fine-grained taint
analysis to find ``hot bytes'', which are the input bytes that flow into
critical system calls or API calls.
A similar idea is presented by other security
researchers~\cite{duran:cansecwest:2011,iozzo:blackhat:2010}.
Dowser~\cite{haller:usec:2013} performs a static analysis during compilation to
find loops that are likely to contain bugs based on a heuristic. Specifically,
it looks for loops containing pointer dereferences. It then computes the
relationship between input bytes and the candidate loops with a taint analysis.
Finally, Dowser runs dynamic symbolic execution while making only the critical
bytes to be symbolic hence improving performance.
VUzzer~\cite{rawat:ndss:2017} and GRT~\cite{ma:ase:2015} leverage both static
and dynamic analysis techniques to extract control- and data-flow features from
the PUT and use them to guide input generation.

Angora~\cite{chen:oakland:2018} and RedQueen~\cite{aschermann:redqueen:2019}
decrease the cost of their analysis by first running for each seed with a costly
instrumentation and using this information for generating inputs which are run
with a lighter instrumentation.
Angora improves upon the ``hot bytes'' idea by using
taint analysis to associate each path constraint to corresponding bytes. It then
performs a search inspired by gradient descent algorithm to guide its mutations
towards solving these constraints.
On the other hand, RedQueen tries to detect how inputs are used in the PUT by
instrumenting all comparisons and looking for correspondence between their
operands and the given input. Once a match is found, it can be used to solve a
constraint.

\subsubsection{PUT Mutation}

One of the practical challenges in fuzzing is bypassing a checksum
validation. For example, when a PUT computes a checksum of an input
before parsing it, many test cases will be rejected by the PUT. To
handle this challenge,
TaintScope~\cite{wang:oakland:2010} proposed a checksum-aware fuzzing
technique, which identifies a checksum test instruction with a
taint analysis, and patches the PUT to bypass the checksum validation.
Once they find a program crash, they generate the correct checksum for
the input to generate a test case that crashes the unmodified PUT.
Caballero \etal~\cite{caballero:ccs:2010} suggested a technique called
stitched dynamic symbolic execution that can generate test cases in
the presence of checksums.

T-Fuzz~\cite{peng:oakland:2018} extends this idea to efficiently penetrate all
kind of conditional branches with grey-box fuzzing.
It first builds a set of Non-Critical Checks (NCC), which are branches that can
be transformed without modifying the program logic.
When the fuzzing campaign stops discovering new paths, it picks an NCC,
transforms it, and then restarts a fuzzing campaign on the modified PUT.
Finally, when a crash is found fuzzing a transformed program, T-Fuzz tries to
reconstruct it on the original program using symbolic execution.

\section{Input Evaluation} \label{sec:inputeval}

After an input is generated, the fuzzer executes the PUT on the input,
and decides what to do with the resulting execution.
This process is called \emph{input evaluation}.
Although the simplicity of executing a PUT is one of the reasons that
fuzzing is attractive, there are many optimizations and design
decisions related to input evaluation that effect the performance and
effectiveness of a fuzzer, and we explore these considerations in this
section.

\subsection{Bug Oracles}

The canonical security policy used with fuzz testing considers every
program execution terminated by a fatal signal (such as a
segmentation fault) to be a violation. This policy detects many memory
vulnerabilities, since a memory vulnerability that overwrites a data
or code pointer with an invalid value will usually cause a
segmentation fault or abort when it is dereferenced.
In addition, this policy is efficient and simple to implement, since
operating systems allow such exceptional situations to be trapped by
the fuzzer without any instrumentation.

However, the traditional policy of detecting crashes will not detect
every memory vulnerability that is triggered. For example, if a stack
buffer overflow overwrites a pointer on the stack with a valid memory
address, the program might run to completion with an invalid result
rather than crashing, and the fuzzer would not detect this. To
mitigate this, researchers have proposed a variety of efficient
program transformations that detect unsafe or unwanted program
behaviors and abort the program. These are often called
\emph{sanitizers}.
\val{Note for future: would be nice to refer to SoK here. One to appear next
year in Oakland and also the "analysis" done on Bug Oracles for embedded system
by Eurocom.}

\subsubsection{Memory and Type Safety}

Memory safety errors can be separated into two classes: spatial and
temporal.  Informally, spatial memory errors occur when a pointer is
dereferenced outside of the object it was intended to point to.  For
example, buffer overflows and underflows are canonical examples of
spatial memory errors.  Temporal memory errors occur when a pointer is
accessed after it is no longer valid.  For example, a use-after-free
vulnerability, in which a pointer is used after the memory it pointed
to has been deallocated, is a typical temporal memory error.

Address Sanitizer (ASan)~\cite{serebryany:atc:2012} is a fast memory
error detector that instruments programs at compile time.  ASan can
detect spatial and temporal memory errors and has an average slowdown
of only 73\%, making it an attractive alternative to a basic crash
harness.  ASan employs a shadow memory that allows each memory address
to be quickly checked for validity before it is dereferenced, which
allows it to detect many (but not all) unsafe memory accesses, even if
they would not crash the original program.
MEDS~\cite{han:ndss:2018} improves on ASan by leveraging the large
memory space available on 64-bit platforms to create large chunks of
inaccessible memory \emph{redzones} in between allocated
objects. These redzones make it more likely that a corrupted pointer
will point to invalid memory and cause a crash.

SoftBound/CETS~\cite{nagarakatte:pldi:2009,nagarakatte:ismm:2010} is another
memory error detector that instruments programs during compilation.
Rather than tracking valid memory addresses like ASan, however,
SoftBound/CETS associates bounds and temporal information with each
pointer, and can theoretically detect all spatial and temporal memory
errors. However, as expected, this completeness comes with a higher
average overhead of 116\%~\cite{nagarakatte:ismm:2010}.
CaVer~\cite{lee:usec:2015}, TypeSan~\cite{haller:sigsac:2016} and
HexType~\cite{jeon:ccs:2017} instrument programs during compilation so that they
can detect \emph{bad-casting} in C++ type casting.  Bad casting occurs when an
object is cast to an incompatible type, such as when an object of a base class
is cast to a derived type.  CaVer has been shown to scale to web browsers, which
have historically contained this type of vulnerability, and imposes between 7.6
and 64.6\% overhead.

Another class of memory safety protection is \emph{Control Flow
  Integrity}~\cite{abadi:ccs:2005,abadi:2009} (CFI), which
detects control flow transitions at runtime that are not
possible in the original program.  CFI can be used to detect test cases
that have illegally modified the control flow of a program.
A recent project focused  on protecting against
 a subset of CFI violations has landed
 in the mainstream \texttt{gcc} and \texttt{clang}
 compilers~\cite{tice:usec:2014}.

\subsubsection{Undefined Behaviors}

Languages such as C contain many behaviors that are left undefined by
the language specification. The compiler is free to handle these
constructs in a variety of ways.  In many cases, a programmer may
(intentionally or otherwise) write their code so that it is only
correct for some compiler implementations.  Although this may not seem
overly dangerous, many factors can impact how a compiler implements
undefined behaviors, including optimization settings, architecture,
compiler, and even compiler version.  Vulnerabilities and bugs often
arise when the compiler's implementation of an undefined behavior does
not match the programmer's expectation~\cite{wang:sosp:2013}.

Memory Sanitizer (MSan) is a tool that instruments programs during
compilation to detect undefined behaviors caused by uses of
uninitialized memory in C and C++~\cite{stepanov:cgo:2015}.  Similar to
ASan, MSan uses a shadow memory that represents whether each
addressable bit is initialized or not.  Memory Sanitizer has
approximately 150\% overhead.
Undefined Behavior Sanitizer (UBSan)~\cite{dietz:icse:2012} modifies
programs at compile-time to detect undefined behaviors.  Unlike other
sanitizers which focus on one particular source of undefined behavior,
UBSan can detect a wide variety of undefined behaviors, such as using
misaligned pointers, division by zero, dereferencing null pointers,
and integer overflow.
Thread Sanitizer (TSan)~\cite{serebryany:wbia:2009} is a compile-time
modification that detects data races with a trade-off between
precision and performance. A data race occurs when two threads
concurrently access a shared memory location and at least one of the
accesses is a write.  Such bugs can cause data corruption and can be
extremely difficult to reproduce due to
non-determinism.

\subsubsection{Input Validation}

Testing for input validation vulnerabilities such as XSS (cross site
scripting) and SQL injection vulnerabilities is a challenging problem,
as it requires understanding the behavior of the very complicated
parsers that power web browsers and database engines.
KameleonFuzz~\cite{duchene:codaspy:2014} detects successful XSS attacks
by parsing test cases with a real web browser, extracting the Document
Object Model tree, and comparing it against manually specified
patterns that indicate a successful XSS attack.
$\mu$4SQLi~\cite{appelt:issta:2014} uses a similar trick to detect SQL
injections.  Because it is not possible to reliably detect SQL
injections from a web applications response, $\mu$4SQLi uses a
database proxy that intercepts communication between the target web
application and the database to detect whether an input triggered
harmful behavior.

\subsubsection{Semantic Difference}

Semantic bugs are often discovered using a technique called
\emph{differential testing}~\cite{mckeeman:1998}, which compares the
behavior of similar (but not identical) programs.
Several
fuzzers~\cite{brubaker:oakland:2014,petsios:oakland:2017,chen:pldi:2016}
have used differential testing to identify discrepancies between
similar programs, which are likely to indicate a bug.
Jung~\etal~\cite{jung:ccs:2008} introduced \emph{black-box
  differential fuzz testing}, which uses differential testing of
multiple inputs on a single program to map mutations from the PUT's
input to its output.  These mappings are used to identify information
leaks.

\subsection{Execution Optimizations}

Our model considers individual fuzz iterations to be executed sequentially.
While the straightforward implementation of such an approach would simply load
the PUT every time a new process is started at the beginning of a fuzz
iteration, the repetitive loading processes can be significantly reduced.
To this end, modern fuzzers provide functionalities that skip over
these repetitive loading processes.
For example, AFL~\cite{aflfuzz} provides a fork-server that allows
each new fuzz iteration to fork from an already initialized process.
Similarly, in-memory fuzzing is another way to optimize the execution
speed as discussed in \S\ref{sssec:inmemory}.
Regardless of the exact mechanism, the overhead of loading and
initializing the PUT is amortized over many iterations.
Xu~\etal~\cite{xu:ccs:2017} further lower the cost of an iteration by designing
a new system call that replaces \texttt{fork()}.

\subsection{Triage}
\label{sec:triage}

Triage is the process of analyzing and reporting test cases that cause
policy violations. Triage can be separated into three steps:
deduplication, prioritization, and test case minimization.

\subsubsection{Deduplication}

\emph{Deduplication} is the process of pruning any test case from the
output set that triggers the same bug as another test case. Ideally,
deduplication would return a set of test cases in which each triggers
a unique bug.

Deduplication is an important component of most fuzzers for
several reasons.  As a practical implementation manner, it avoids
wasting disk space
and other resources by storing duplicate results on
the hard drive.  As a usability consideration, deduplication makes it
easy for users to understand roughly how many different bugs are
present, and to be able to analyze an example of each bug.  This is
useful for a variety of fuzzer users; for example, attackers
may want to look only for ``home run'' vulnerabilities that are likely
to lead to reliable exploitation.

There are currently three major deduplication implementations used in
practice: stack backtrace hashing, coverage-based deduplication, and
semantics-aware deduplication.

\begin{figure}
  \tikzstyle{freecell}=[draw=black]
\tikzstyle{occupiedcell}=[fill=blue!10!orange!10,draw=black]

\begin{tikzpicture}[scale=15cm/\textwidth]
  \cell {main}
  \startframe
  \cell {d}
  \cell {c}
  \cell {b}
  \cell {a}
  \bcell {foo (crashed~\faBolt)}
  \finishframe {$n=5$}
\end{tikzpicture}
  \caption{Stack backtrace hashing example.}
  \label{fig:stack}
\end{figure}
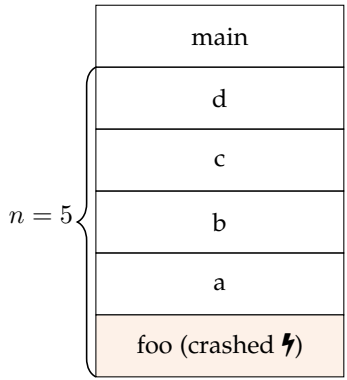

\paragraph{Stack Backtrace Hashing}
Stack backtrace hashing~\cite{molnar:2009} is one of the oldest and
most widely used methods for deduplicating crashes, in which an
automated tool records a stack backtrace at the time of the crash, and
assigns a \emph{stack hash} based on the contents of that backtrace.
For example, if the program crashed while executing a line of code in
function \texttt{foo}, and had the call stack
$\texttt{main} \rightarrow \texttt{d} \rightarrow \texttt{c}
\rightarrow \texttt{b} \rightarrow \texttt{a} \rightarrow
\texttt{foo}$ (see Fig.~\ref{fig:stack}), then a stack backtrace hashing
implementation with $n=5$ would group that test case with other
crashing executions whose backtrace ended with
$\texttt{d} \rightarrow \texttt{c} \rightarrow \texttt{b} \rightarrow
\texttt{a} \rightarrow \texttt{foo}$.
Stack hashing implementations vary widely, starting with the number of
stack frames that are included in the hash.  Some implementations use
one~\cite{crashwrangler}, three~\cite{molnar:2009,woo:ccs:2013},
five~\cite{gdbexploitable,bff-deprecated-merged-with-certfuzz}, or do not have any
limit~\cite{kargen:fse:2015}.  Implementations also differ in the amount
of information included from each stack frame.  Some implementations
will only hash the function's name or address, but other
implementations will hash both the name and the offset or line.
Neither option works well all the time, so some
implementations~\cite{exploitable, gdbexploitable} produce two hashes:
a major and minor hash.  The major hash is likely to group dissimilar
crashes together as it only hashes the function name, whereas the
minor hash is more precise since it uses the function name and line
number, and also includes an unlimited number of stack frames.

Although stack backtrace hashing is widely used, it is not without its shortcomings.
The underlying hypothesis of stack backtrace hashing is that similar
crashes are caused by similar bugs, and vice versa, but, to the best of our knowledge, this
hypothesis has never been directly tested.
There is some reason to doubt its veracity: some crashes
do not occur near the code that caused the crash.  For example, a
vulnerability that causes heap corruption might only crash when an
unrelated part of the code attempts to allocate memory, rather than
when the heap overflow occurred.

\paragraph{Coverage-based Deduplication}
AFL~\cite{aflfuzz} is a popular grey-box fuzzer that employs an
efficient source-code instrumentation to record the edge coverage of
each execution of the PUT, and also measure coarse hit counts for each
edge.  As a grey-box fuzzer, AFL primarily uses this coverage
information to select new seed files.  However, it also leads to a
fairly unique deduplication scheme as well.  As described by its
documentation, AFL considers a crash to be unique if either
\begin{enumerate*}[(i)]
\item the crash covered a previously unseen edge, or
\item the crash did \emph{not} cover an edge that was present in all
  earlier crashes.
\end{enumerate*}

\paragraph{Semantics-aware Deduplication}

RETracer \cite{cui:icse:2016} performs crash triage based on the
semantics recovered from a reverse data-flow analysis from each
crash. Specifically, RETracer checks which pointer caused the crash by
analyzing a crash dump (core dump), and recursively identifies which
instruction assigned the bad value to it. It eventually finds a
function that has the maximum frame level\maeg{undefined term}, and
``blames'' the function. The blamed function can be used to cluster
crashes. The authors showed that their technique successfully deduped
millions of Internet Explorer bugs into one.  In contrast, stack hashing
categorized the same crashes into a large number of different groups.

\subsubsection{Prioritization and Exploitability}

Prioritization, \aka the fuzzer taming problem~\cite{chen:pldi:2013}, is
the process of ranking or grouping violating test cases according to
their severity and uniqueness.  Fuzzing has traditionally been
used to discover memory vulnerabilities, and in this context
prioritization is better known as determining the
\emph{exploitability} of a crash.  Exploitability informally describes
the likelihood of an adversary being able to write a practical exploit
for the vulnerability exposed by the test case.  Both defenders and
attackers are interested in exploitable bugs.  Defenders generally fix
exploitable bugs before non-exploitable ones, and attackers are
interested in exploitable bugs for obvious reasons.

One of the first exploitability ranking systems was Microsoft's
!exploitable~\cite{exploitable}, which gets its name from the
\texttt{!exploitable} WinDbg command name that it provides.  %
!exploitable employs several heuristics paired with a simplified taint
analysis~\cite{newsome:ndss:2005,schwartz:oakland:2010}.
It classifies each crash on the following severity scale: \texttt{EXPLOITABLE}
$>$ \texttt{PROBABLY\_EXPLOITABLE} $>$ \texttt{UNKNOWN} $>$
\texttt{NOT\_LIKELY\_EXPLOITABLE}, in which $x > y$ means that $x$ is more
severe than $y$.  Although these classifications are not formally defined,
!exploitable is informally intended to be conservative and error on the side of
reporting something as more exploitable than it is.  For example, !exploitable
concludes that a crash is \texttt{EXPLOITABLE} if an illegal instruction is
executed, based on the assumption that the attacker was able to coerce control
flow.  On the other hand, a division by zero crash is considered
\texttt{NOT\_LIKELY\_EXPLOITABLE}.

Since !exploitable was introduced, other, similar rule-based heuristics systems
have been proposed, including the exploitable plugin for
GDB~\cite{gdbexploitable} and Apple's
CrashWrangler~\cite{crashwrangler}. However, their correctness has not been
systematically studied and evaluated yet.

\subsubsection{Test case minimization}
Another important part of triage is \emph{test case minimization}.
Test case minimization is the process of identifying the portion of a
violating test case that is necessary to trigger the violation, and
optionally producing a test case that is smaller and simpler than the
original, but still causes a violation.
Although test case minimization and seed trimming
(see~\ref{sec:seed-trimming}, p. \pageref{sec:seed-trimming}) are
conceptually similar in that they aim at reducing the size of an input, they
are distinct because a minimizer can leverage a bug oracle.

Some fuzzers use their own implementation and algorithms for this.
BFF~\cite{bff-deprecated-merged-with-certfuzz} includes a minimization algorithm tailored to
fuzzing~\cite{householder:minimization:2012} that attempts to minimize the number
of bits that are different from the original seed file.
AFL~\cite{aflfuzz} also includes a test case minimizer, which attempts
to simplify the test case by opportunistically setting bytes to zero
and shortening the length of the test case.  Lithium~\cite{lithium} is
a general purpose test case minimization tool that minimizes files by
attempting to remove ``chunks'' of adjacent lines or bytes in
exponentially descending sizes.  Lithium was motivated by the
complicated test cases produced by JavaScript fuzzers such as
jsfunfuzz~\cite{funfuzz}.

There are also a variety of test case reducers that are not
specifically designed for fuzzing, but can nevertheless be used for
test cases identified by fuzzing.  These include format agnostic
techniques such as delta debugging~\cite{zeller:tse:2002}, and specialized
techniques for specific formats such as C-Reduce~\cite{regehr:pldi:2012}
for C/C++ files.  Although specialized techniques are obviously
limited in the types of files they can reduce, they have the advantage
that they can be significantly more efficient than generic techniques,
since they have an understanding of the grammar they are trying to
simplify.

\section{Configuration Updating}\label{sec:confupdate}

The \confUpdate function plays a critical role in distinguishing the
behavior of black-box fuzzers from grey- and white-box fuzzers. As
discussed in Algorithm~\ref{algo:fuzzer}, the \confUpdate function can
modify the set of configurations ($\confs$) based on the configuration
and execution information collected during the current fuzzing run.
In its simplest form, \confUpdate returns the $\confs$ parameter unmodified.
Black-box fuzzers do not perform any program introspection beyond evaluating the
bug oracle \bugoracle, and so they typically leave \confs unmodified because
they do not have any information collected that would allow them to modify
it\footnote{Some fuzzers add violating test cases to the set of seeds.  For
example, BFF~\cite{bff-deprecated-merged-with-certfuzz} calls this feature crash
recycling.}.

In contrast, grey- and white-box fuzzers are distinguished by their
more sophisticated implementations of the \confUpdate function, which
allows them to incorporate new fuzz configurations, or remove old ones
that may have been superseded.
\confUpdate enables information collected during
one fuzzing iteration to be used by all future fuzzing iterations.
For example, white-box fuzzers typically create a new fuzz
configuration for every new test case produced, since they produce
relatively few test cases compared to black- and grey-box fuzzers.

\subsection{Evolutionary Seed Pool Update} \label{ssec:evolutionary}

An Evolutionary Algorithm (EA) is a heuristic-based approach that
involves biological evolution mechanisms such as mutation,
recombination, and selection.
In the context of fuzzing, an EA maintains a \emph{seed pool} of
promising individuals (i.e., seeds) that evolves over the course of a
fuzzing campaign as new individuals are discovered.
Although the concept of EAs is relatively simple, it forms the basis
of many grey-box fuzzers~\cite{aflfuzz, libfuzzer, syzkaller}.
The process of choosing the seeds to be mutated and the mutation
process itself were detailed in \S\ref{sec:fcs-greybox} and
\S\ref{sec:inputgen} respectively.

Arguably, the most important step of an EA is to add a new
configuration to the set of configurations \confs, which occurs during
the \confUpdate step of fuzzing.
Most EA-based fuzzers use node or branch coverage as a fitness
function: if a new node or branch is discovered by a test case, it is
added to the seed pool.
As the number of reachable paths can be orders of magnitude larger
than the number of seeds, the seed pool is intended to be a
\emph{diverse} subselection of all reachable paths in order to
represent the current exploration of the PUT.
Also note that seed pools of different size can have the same coverage
(as mentioned in~\S\ref{ssec:seedselection},
p.~\pageref{ssec:seedselection}).

A common strategy in EA fuzzers is to refine the fitness function so
that it can detect more subtle and granular indicators of
improvements.
For example, AFL~\cite{aflfuzz} refines its fitness function
definition by recording the number of times a branch has been taken.
STADS~\cite{bohme:tosem:2018} presents a statistical framework
inspired by ecology to estimate how many new configurations will be
discovered if the fuzzing campaign continues.
Another common strategy is to measure the fraction of conditions that are met
when complex branch conditions are evaluated.
For example, \texttt{LAF-INTEL}~\cite{lafintel:afl} simply breaks
multi-byte comparison into several branches, which allows it to detect
when a new seed passes an intermediate byte comparison.
LibFuzzer~\cite{libfuzzer}, Honggfuzz~\cite{honggfuzz}, go-fuzz~\cite{gofuzz}
and Steelix~\cite{li:fse:2017} instrument all comparisons, and add any test case
that makes progress on a comparison to the seed pool.
A similar idea was also released as a stand-alone instrumentation module
for \texttt{clang}~\cite{cmpcov}.
Additionally, Steelix~\cite{li:fse:2017} checks which input offsets
influence comparison instructions.
Angora~\cite{peng:oakland:2018} improves the fitness criteria of AFL
by considering the calling context of each branch taken.

VUzzer~\cite{rawat:ndss:2017} is an EA-based fuzzer whose fitness
function relies on the results of a custom program analysis that
determines weights for each basic block.
Specifically, VUzzer first uses a built-in program analysis to
classify basic blocks as either normal or error handling (EH).
For a normal block, its weight is inversely proportional to the
probability that a random walk on the CFG containing this block visits
it according to transition probabilities defined by VUzzer.
This encourages VUzzer to prefer configurations that exercise normal
blocks deemed rare by the aforementioned random walk.
The weight of EH blocks is \emph{negative}, and its magnitude is the ratio of the number of
basic blocks compared to the number of EH blocks exercised by this configuration.
These negative weights are used to discourage the execution of error
handling (EH) blocks, based on the hypothesis that traversing an EH
block signals a lower chance of exercising a vulnerability since bugs
often coincide with unhandled errors.

\subsection{Maintaining a Minset} \label{ssec:culling}

With the ability to create new fuzzing configurations comes the risk
of creating too many configurations. A common strategy used to
mitigate this risk is to maintain a \emph{minset}, or a minimal set of
test cases that maximizes a coverage metric. Minsetting is also used
during \preprocess, and is described in more detail in
\S\ref{ssec:seedselection}.

Some fuzzers use a variant of maintaining a minset that is specialized
for configuration updates.
As one example, rather than completely removing configurations that are not in
the minset, which is what Cyberdyne~\cite{goodman:oakland:2018} does,
AFL~\cite{aflfuzz} uses a \emph{culling} procedure to mark minset configurations
as being \emph{favorable}.
Favorable fuzzing configurations are given a significantly higher chance of
being selected for fuzzing by the \schedule function.  The author of AFL notes
that ``this provides a reasonable balance between queue cycling speed and test
case diversity''~\cite{afltech}.

\section{Related Work} \label{s:related}

The literature on fuzzing had an early bloom in 2007--2008, when three
trade-books on the subject were published within the two-year
period~\cite{evron:2007, sutton:2007, takanen:2008}.
These books took a more practical approach by presenting the different tools and
techniques available at the time and their usages on a variety of targets.
We note that Takanen~\etal~\cite{takanen:2008} already distinguished among
black-, white- and grey-box fuzzers, although no formal definitions were given.
\val{Their def rely on environmental criterias like "access to the source".}
Most recently, \cite{takanen:2008} had been revised after a decade.
The second edition~\cite{takanen:2018} contained many updates to include modern
tools such as AFL~\cite{aflfuzz} and ClusterFuzz~\cite{clusterfuzz}.

We are aware of two fuzzing surveys that are concurrent to
  ours~\cite{li:cybsec:2018, liang:tr:2018}.
However, the goals of both of these surveys are more focused than ours, which is
to provide a comprehensive study on recent developments covering the entire
area.
In particular, Li~\etal~\cite{li:cybsec:2018} provided a thorough review of many
recent advances in fuzzing, though the authors have also decided to focus on the
detail of coverage-based fuzzing and not others.
More similar to ours, Liang~\etal\cite{liang:tr:2018} proposed an informal model
to describe various fuzzing techniques.
However, their model is not flexible enough to encompass some of the fuzzing
approaches we cover in this paper, such as model inference
(see~\S\ref{sssec:infmodel}) and hybrid fuzzing (see~\S\ref{ssec:whitebox}).

Klees~\etal~\cite{klees:ccs:2018} recently found that there has been no coherent
way of evaluating fuzzing techniques, which can hamper our ability to compare
the effectiveness of fuzzing techniques.
In addition, they provided several useful guidelines for evaluating fuzzing
algorithms.
We consider their work to be orthogonal to ours as the evaluation of fuzzing
algorithms is beyond the scope of this paper.

\section{Concluding Remarks}\label{sec:conclusion}

As we have set forth in \S\ref{sec:intro}, our goal for this paper is to
distill a comprehensive and coherent view of modern fuzzing literature.
To this end, we first present a general-purpose model fuzzer to facilitate our
effort to explain the many forms of fuzzing in current use.
Then, we illustrate a rich taxonomy of fuzzers using Figure~\ref{fig:genealogy}
(p.~\pageref{fig:genealogy}) and Table~\ref{tab:overview}
(p.~\pageref{tab:overview}).
We have explored every stage of our model fuzzer by discussing the design
decisions as well as showcasing the many achievements by the community at large.

{
    \raggedright
    \sloppy
    \IEEEtriggeratref{237} %
    \bibliographystyle{IEEEtranS}
    \bibliography{bibliography,misc}
}

\end{document}